\documentclass[aps,twocolumn,english,superscriptaddress,citeautoscript,preprintnumbers,amsmath,amssymb,floatfix,footinbib]{revtex4-1}
\usepackage{graphicx}
\usepackage{graphics}               
\usepackage{bm}   
\usepackage{amssymb } 
\usepackage{array}       
\usepackage[breaklinks=true,colorlinks,citecolor=blue,linkcolor=blue,urlcolor=blue]{hyperref}
\usepackage{float}
\usepackage[table]{xcolor}

\begin{document}
\title{Magnetotransport properties of the topological nodal-line semimetal CaCdSn}

\author{Antu Laha}
\affiliation{Department of Physics, Indian Institute of Technology, Kanpur 208016, India}

\author{Sougata Mardanya}
\affiliation{Department of Physics, Indian Institute of Technology, Kanpur 208016, India}

\author{Bahadur Singh}
\email{bahadursingh24@gmail.com}
\affiliation{Department of Physics, Northeastern University, Boston, Massachusetts 02115, USA}

\author{Hsin Lin}
\affiliation{Institute of Physics, Academia Sinica, Taipei 11529, Taiwan}

\author{Arun Bansil}
\affiliation{Department of Physics, Northeastern University, Boston, Massachusetts 02115, USA}

\author{Amit Agarwal}
\email{amitag@iitk.ac.in}
\affiliation{Department of Physics, Indian Institute of Technology, Kanpur 208016, India}

\author{Z. Hossain}
\email{zakir@iitk.ac.in}
\affiliation{Department of Physics, Indian Institute of Technology, Kanpur 208016, India}

\begin{abstract}
Topological nodal-line semimetals support protected band crossings which form nodal lines or nodal loops between the valence and conduction bands and exhibit novel transport phenomena. Here we address the topological state of the nodal-line semimetal candidate material, CaCdSn, and report magnetotransport properties of its single crystals grown by the self-flux method. Our first-principles calculations show that the electronic structure of CaCdSn harbors a single nodal loop around the $\Gamma$ point in the absence of spin-orbit coupling (SOC) effects. The nodal crossings in CaCdSn are found to lie above the Fermi level and yield a Fermi surface that consists of both electron and hole pockets. CaCdSn exhibits high mobility ($\mu \approx 3.44\times 10^4$ cm$^2$V$^{-1}$s$^{-1}$) and displays a field-induced metal-semiconductor like crossover with a plateau in resistivity at low temperature. We observe an extremely large and quasilinear non-saturating transverse as well as longitudinal magnetoresistance (MR) at low temperatures ($\approx 7.44\times 10^3 \%$ and $\approx 1.71\times 10^3\%$, respectively, at 4K). We also briefly discuss possible reasons behind such a large quasilinear magnetoresistance and its connection with the nontrivial band structure of CaCdSn.
\end{abstract}

\maketitle

\section{Introduction}

Topological semimetals are drawing intense current interest as platforms for transformational fundamental science studies as well as their potential for next-generation technological applications \cite{bansil2016,RevModPhys.82.3045,RevModPhys.90.015001}. These semimetals exhibit symmetry-protected band crossings between the valence and conduction states and present unique responses to applied electric and magnetic fields. Their transport properties include a high carrier mobility, an extremely large and linear magnetoresistance (MR), and a chiral-anomaly-induced negative longitudinal MR \cite{TaAs_Nature_comn_2016,Na3Bi_Science_2015,NbP,ZrTe5_Nature_phy_2018}. Depending on their band crossings, topological semimetals are classified into Dirac, Weyl, and nodal-line semimetals. While the Dirac (Weyl) semimetals support a discrete number of degenerate zero-dimensional (0D) $k$ points in the Brillouin zone (BZ), nodal-line semimetals (NLSMs) possess extended band-touching points that result in the formation of protected one-dimensional (1D) lines or loops in the BZ \cite{fang2016,bian2016,PtSn4_natphy_2016,chang2017,CaAgAs_PRB_R_2017,ZrSiS_PNAS_2017,ZrSiS_PRL_2018,ZrSiS_ScienceAdv_2016}. Despite the experimental realization of a variety of topological semimetals \cite{TaAs_Nature_comn_2016,Na3Bi_Science_2015,NbP,ZrTe5_Nature_phy_2018,fang2016,TaAs_Nature_comn_2016,Na3Bi_Science_2015,NbP,ZrTe5_Nature_phy_2018,PbTaSe2_PRB_2016,PtSn4_natphy_2016,ZrSiS_PRB_R_2016,ZrSiX_PRB_R_2017,CaAgAs_PRB_R_2017}, the choice of currently available materials remains limited. There is thus an urgent need to find new topological semimetals and explore their unique properties.

In this connection, magnetotransport experiments have emerged as a powerful probe for obtaining a handle on the nature of electronic states in topological materials \cite{MoAs2_PRB_2018,Cd3As2_PRL_2015,NdSb_PRB_2016,TlBiSSe_PRB_2015,YSb_SR_2016,CaCdGe_PRB_2017,YbCdGe_PRB_2019,Hu2008}. An extremely large MR has been reported in topological materials at high magnetic fields and low temperatures \cite{Leahy10570,Na3Bi_Science_2015,NbP_natphy_2015,PtBi2_PRL_2017,PtSn4_PRB_2018,NbAs2TaAs2_PRB(R)_2016,NbAs2TaAs2_PRB_2016,Cd3As2_natmat_2015,WTe2,WP2_MoP2_natcom_2017}. The physical origin of such a large non-saturating and linear MR include: (i) semiclassical transport in charge-compensated metals and semimetals \cite{SPAIN197623,CaCdGe_PRB_2017,WTe2_PRB_2015}; (ii) disorder-induced mobility fluctuations\cite{Parish2003}; (iii) unusual electron trajectories in weakly disordered systems \cite{PhysRevB.92.180204}; and (iv) magneto-transport in the extreme quantum limit in the presence of small Fermi surface pockets \cite{Abrikosov_2000,Abrikosov_PRB_1998,Abrikosov_PRB_1999,CrAs_Nature_Comn_2017}. This clearly indicates that material specific analysis of the physical and electronic properties of materials is required to understand the observation of extremely large MR. 

In this paper, we discuss topological electronic state and magnetotransport properties of the topological nodal-line semimetal candidate material CaCdSn. Our systematic first-principles calculations demonstrate that CaCdSn is a nodal-line semimetal that supports a single nodal-loop centered around the $\Gamma$ point in the BZ. Interestingly, the nodal lines are located above the Fermi level and yield a large hole pocket and two small electrons pockets at the Fermi level. Our magnetotransport measurements were carried out on high-quality single crystals under low-temperature/high-field conditions and reveal that CaCdSn possesses an extremely large non-saturating MR ($> 10^3\%$). We also observe a hole-dominated linear Hall resistivity, a quasilinear non-saturating MR, and the absence of quantum oscillations. We discuss a number of possible mechanisms for realizing the large quasilinear MR in CaCdSn. 

An outline of this paper is as follows. We give details of experimental and theoretical methodologies in Sec. \ref{methods}. In Sec. \ref{bandstr}, we explain the topological state and electronic properties of CaCdSn. Sections \ref{exp1} and \ref{exp2} give experimental analysis of the electrical and magnetotransport properties of CaCdSn. Discussion and conclusions are presented in Sec. \ref{disc}.

\section{Methods}\label{methods}

High quality single crystals of CaCdSn were synthesized using cadmium flux. Ca (99.99$\%$, Alfa Aesar), Cd (99.99$\%$, Alfa Aesar) and Sn (99.99$\%$, Alfa Aesar) shots in molar ratio of 1:47:1 were put in an alumina crucible before the crucible was sealed into a quartz tube under partial pressure of argon gas. The quartz tube was heated to 1000$^\circ$C, kept for 3 hours at this temperature, and then cooled to 500$^\circ$C at a rate of 3$^\circ$C/hour \cite{CaCdGe_PRB_2017,YbCdGe_PRB_2019}. Needle-like single crystals were extracted from the flux by centrifuging. The crystal structure was determined by powder x-ray diffraction (XRD) using Cu-K$_\alpha$ radiation in a PANalytical X$'$Pert PRO diffractometer. The phase purity and composition were confirmed via energy-dispersive x-ray spectroscopy (EDS) measurements in a JEOL JSM-6010LA scanning electron microscope. Magnetotransport measurements were carried out in a physical property measurement system (PPMS, Quantum Design) via standard four-probe method.

\begin{figure}
\centering
\includegraphics[width=0.99\linewidth]{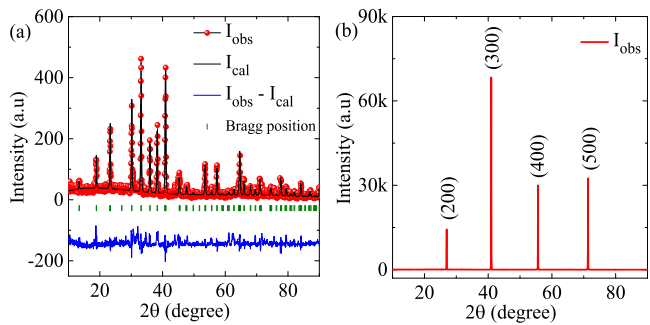}
\caption{(a) Powder x-ray diffraction pattern of crushed CaCdSn single crystals, recorded at room temperature. The observed intensity (red scattered points), Rietveld refinement fit (solid black line), difference between the experimentally observed and calculated intensities (solid blue line) and Bragg peak positions (vertical green bars) are shown. (b) Single crystal x-ray diffraction pattern.
\label{XRD}}
\end{figure}

Rietveld refinement of the XRD data yielded a non-centrosymmetric hexagonal crystal structure with space group $P-62m$ (No. 189) and lattice parameters: $a=b=7.6273(4)$ \AA $~ $and $c=4.7002(8)$ \AA. Notably, the Rietveld structural refinement of the power XRD was done using the FULLPROF software package \cite{Fullprof_1993} as shown in [Fig.~\ref{XRD}(a)]. No impurity phase was observed within our experimental resolution. Crystal axis was determined by single crystal XRD measurement [Fig.~\ref{XRD}(b)]. The EDS analysis confirms the single-phase nature and almost perfect stoichiometry of the grown crystals. 

First-principles calculations were performed within the framework of the density functional theory (DFT) \cite{hohenberg1964,kohn1965} with the projector augmented wave (PAW) \cite{blochl1994_PAW} pseudopotentials, as implemented in the Vienna ab-initio simulation package (VASP) \cite{kresse1996,kresse1999}. The generalized gradient approximation (GGA) with the Perdew-Burke-Ernzerhof parametrization was used to include exchange-correlation effects. The SOC was considered self-consistently to include relativistic effects \cite{perdew1996_gga}.

% XRD

%% Crystal structure and Hall Resistivity 
\begin{figure}
\centering
\includegraphics[width=0.99\linewidth]{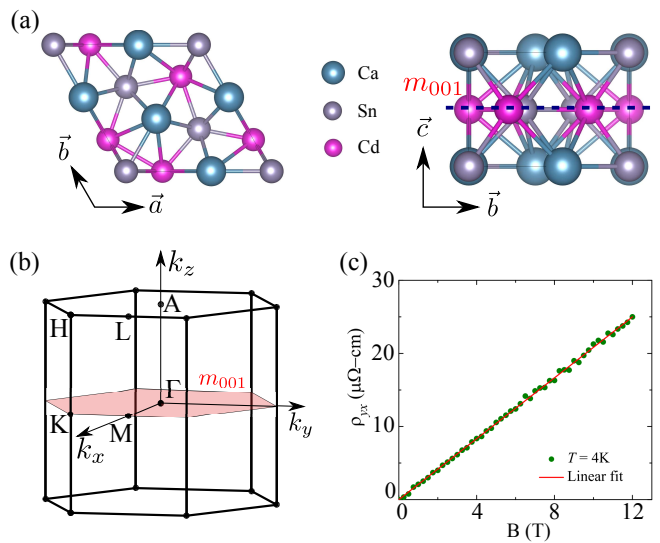}
\caption{(a) Crystal structure of CaCdSn. The blue dashed line in the right panel represents the $m_{001}$ mirror plane. (b) The bulk Brillouin zone with high-symmetry points and $m_{001}$ mirror plane. (c) Hall resistivity as a function of the magnetic field at $T$=4K ($B \perp I$ and $I || c$-axis). 
\label{fig1}}
\end{figure}

%% Band structure and Fermi surface
\begin{figure*}
	\centering
	\includegraphics[width=0.9\linewidth]{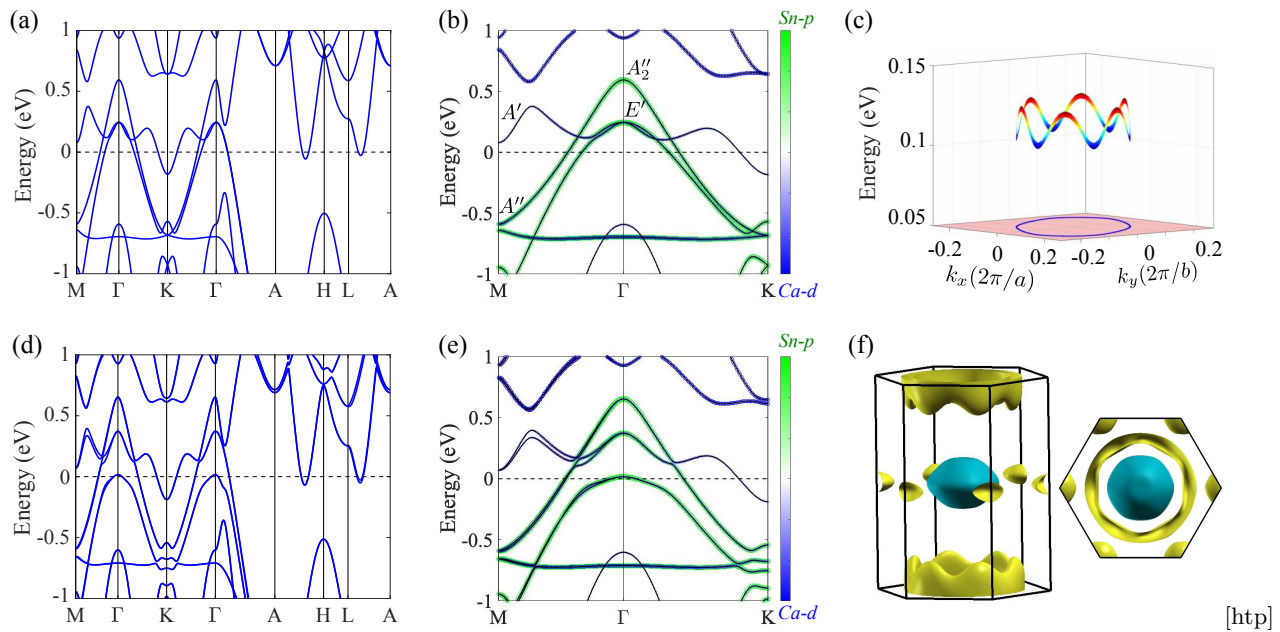}[htp]
	\caption{Nodal-line semimetal state of CaCdSn. (a) Bulk band structure without spin-orbit coupling (SOC) along the various high-symmetry directions in the Brillouin zone (BZ). (b) A closeup of the bands in the $k_z=0$ plane. The color and size of various markers denote the orbital contribution of $Sn-p$ and $Ca-d$ states to various bands. (c) Energy-momentum dispersion of the nodal loop on the $k_x-k_y$ plane ($E-k_x-k_y$). (d)-(e) Same as (a)-(b) but with the inclusion of SOC. (f) Fermi surface with the electron (yellow) and hole (green) pockets.
		\label{fig2}}
\end{figure*}

\section{Electronic structure}\label{bandstr}

Crystal structure of CaCdSn (see Fig.~\ref{fig1}(a)) consists of nine atoms per unit cell in which Ca and Cd occupy Wyckoff positions $3f$ and $3g$, respectively, and Sn is located at Wyckoff positions $2d$ and $1a$. Crystal symmetries include a three-fold rotational axis ($C_{3z}$) and a horizontal mirror plane $m_{001}$ similar to the other known nodal-line semimetals CaAgAs, CaCdGe and YbCdGe\cite{CaAgAs_PRB_R_2017,CaCdGe_PRB_2017,YbCdGe_PRB_2019}. Additionally, CaCdSn respects time-reversal symmetry $\mathcal{T}$ but breaks inversion symmetry $\mathcal{I}$.  
 
Bulk band structure of CaCdSn along the high-symmetry directions without spin-orbit coupling (SOC) is shown in Fig.~\ref{fig2}(a). It is readily seen that both the valence and conduction bands cross the Fermi level, indicating the semimetallic nature of CaCdSn. Interestingly, the valence and conduction bands cross each other on the $m_{001}$ mirror plane at 0.13 eV along the $\Gamma-M$ line and at 0.11 eV along the $\Gamma-K$ direction [see Fig. \ref{fig2}(b)]. These crossing bands which are primarily composed of Sn $p$ and Ca $d$ states, present an inverted band structure at the $\Gamma$ point and belong to $E'$ and $A_2''$ irreducible representations (IRR) with $m_{001}$ mirror eigenvalues $+1$ and $-1$, respectively. The opposite mirror eigenvalues make these band crossings stable and enforce the formation of a nodal loop on the $k_z=0$ plane as shown in Fig.~\ref{fig2}(c). With the inclusion of SOC in the computations, the band hybridization opens a local bandgap at the crossing points [Figs.~\ref{fig2}(d) and \ref{fig2}(e)] and the system makes a transition to a nontrivial state with the $\mathbb{Z}_2$ topological indices $(\nu_0;\nu_1\nu_2\nu_3) = (1;000)$. The nonzero value of the strong topological index ($\nu_0$) is a clear manifestation of nontrivial topology in this system \cite{fu2007topological}.

Although CaCdSn is topologically nontrivial with an inverted band structure around the $\Gamma$ point, it supports electron and hole pockets at the Fermi level as illustrated in Fig.~\ref{fig2}(f). We find three pockets including a large hole pocket around $\Gamma$ and two electron pockets around $K$ and in the $k_z = \pi/c$ plane. The $k_z = \pi/c$ is hollow from the center as shown in the right panel of Fig. \ref{fig2}(f). These Fermi pockets which govern magnetotransport, provide a fermiology in CaCdSn that is different from that of its iso-structural cousins CaAgAs and CaCdGe that host an equal number of electron and hole pockets \cite{CaCdGe_PRB_2017}. 

\section{Mobility and resistivity}\label{exp1} 

We now turn to discuss our experimental results on CaCdSn.  To determine the charge-carrier density and mobility, we measured the magnetic field dependent Hall resistivity ($\rho_{yx}$) at 4K [see Fig.~\ref{fig1}(c)].  The MR contribution can be removed by symmetrizing the Hall data: $\rho_{yx}=[\rho_{yx}(B)-\rho_{yx}(-B)]/2$. The positive and linear $\rho_{yx}$ indicates that the hole carriers dominate in CaCdSn. This result is somewhat surprising, since our calculated Fermi surface contains both electron and hole pockets. The linear fit to the $\rho_{yx}$ data yields a Hall coefficient of $R_H = 2.1$ $\mu \Omega$-cm T$^{-1}$. The density of the hole carriers is estimated to be $n_h=1/(e R_H) = 2.98 \times 10^{20}$ cm$^{-3}$, and the Hall mobility to be $\mu_h =R_H/\rho_{(B=0)} = 3.44\times 10^4$ cm$^2$V$^{-1}$s$^{-1}$. Although this carrier density is significantly higher than that of typical Dirac/Weyl semimetals ($n \approx 10^{17}-10^{18}$ cm$^{-3}$) \cite{Cd3As2_PRL_2015,TaAs_PRX}, it is of the same order of magnitude as that reported in a number of nodal-line semimetals \cite{CaCdGe_PRB_2017,YbCdGe_PRB_2019,NLSM_ZrSiS,NLSM_ZrSiSe_ZrSiTe}.

%% Resistivity
\begin{figure}[htp]
\centering
\includegraphics[width=0.99\linewidth]{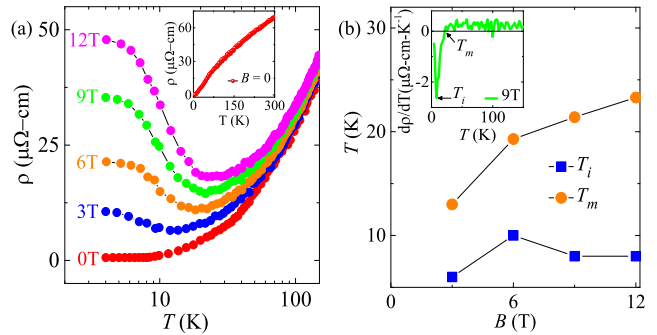} 
\caption{(a) Resistivity as a function of temperature for magnetic field up to 12T. (b) $T_m$ and $T_i$ as a function of $B$; inset shows the first derivative of resistivity with temperature as a function of $T$ for $B$=9T. ($B\perp I $ and $I||c$-axis).
\label{fig3}}
\end{figure}

The measured resistivity of CaCdSn as a function of temperature for various magnetic field values is shown in Fig.~\ref{fig3}(a). The resistivity decreases as we lower the temperature and finally shows a saturation below 8K with a low residual resistivity of $\rho = 0.6~\mu\Omega$-cm (see inset of Fig.~\ref{fig3}(a)). This leads to a high residual resistivity ratio (RRR) of 114, which indicates the high quality of our single crystals. In the presence of a magnetic field, the $\rho-T$ curves in Fig.~\ref{fig3}(a) show a metal-semiconductor-like crossover which becomes more noticeable as $B$ is increased from 3T to 12T. These features are observed in many topological semimetals \cite{PhysRevLett.94.166601, YbCdGe_PRB_2019,LaSb_NaturePhy_2015,NbAs2TaAs2_PRB(R)_2016,TaSb2_PRB(R)_2016,LaSb_PRL_2016,PtBi2_PRL_2017,ZrSiS_ScienceAdv_2016,LaSbTe_pPRB_2017,ZrSiS_PNAS_2017, du2018}. The upturn in the resistivity at low temperature is a consequence of field induced corrections to resistivity in topological materials,  which become more prominent at high magnetic field \cite{WTe2_PRB_2015}. 

From the first derivative of $\rho(T)$, we obtain two characteristic temperatures - the minimum of the $d\rho/dT$ curve ($T_i$) and the temperature associated with the crossover point $d{\rho}/dT = 0$ ($T_m$) as noted in the inset of Fig.~\ref{fig3}(b). Here, $T_m$ marks the temperature associated with the metal-semiconductor-like crossover whereas $T_i$ corresponds to the inflection point in resistivity where $d\rho/dT$ is minimum. The residual resistivity plateaus appear just below $T_i$. The variation of $T_m$ and $T_i$ with magnetic field is shown in Fig.~\ref{fig3}(b).

\section{Magnetoresistance} \label{exp2} 

\begin{figure}[htp]
\centering
\includegraphics[width=0.9\linewidth]{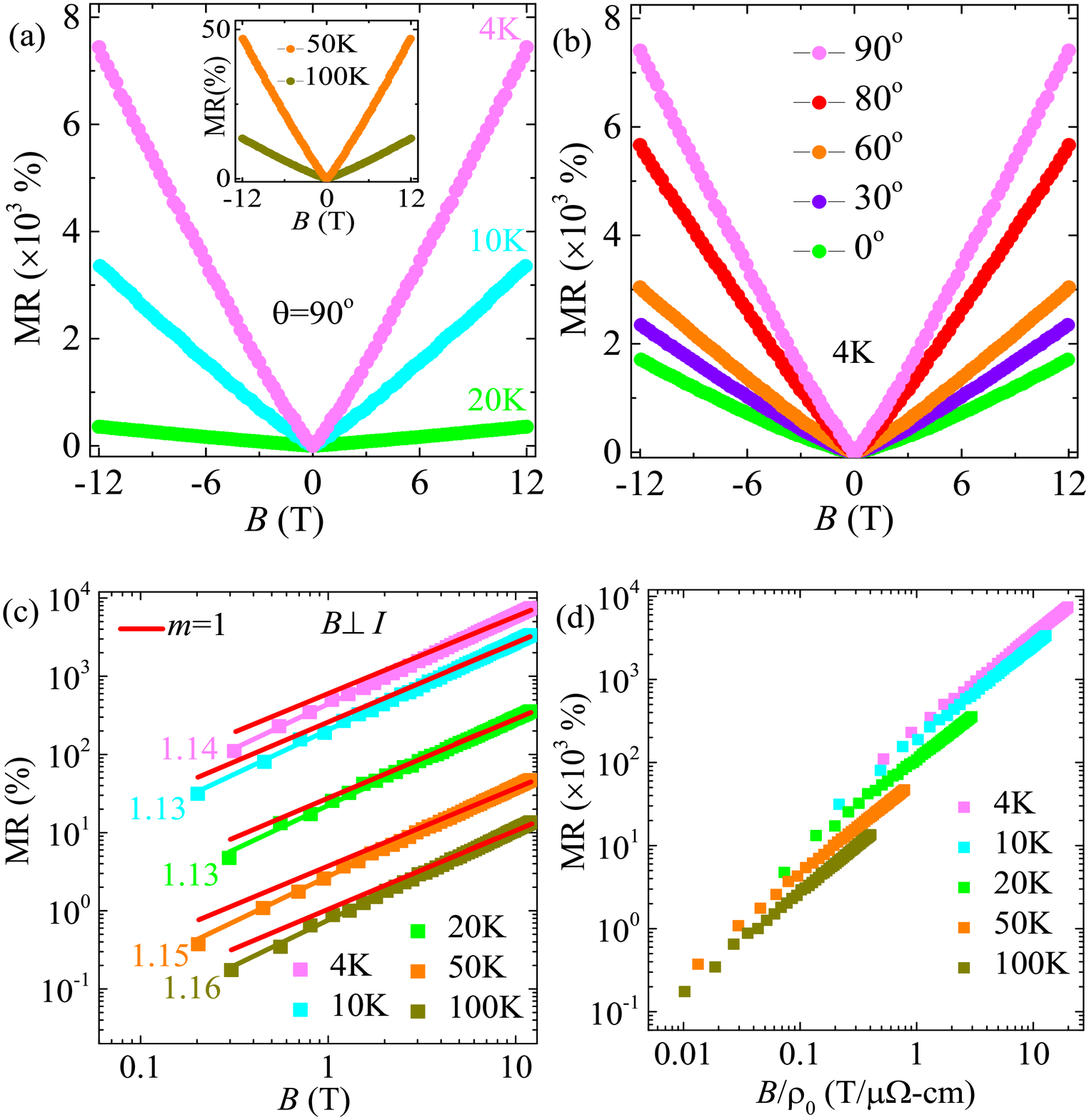}
\caption{(a) Magnetic field dependent magnetoresistance (MR) for CaCdSn at $T = 4K$, $10K$, and $20K$ ($B\perp I$ and $I||c$-axis). The inset shows MR at $50K$ and $100K$. (b) MR as a function of the applied magnetic field at various angles $\theta$ between $B$ and $I$ at $4K$. The presence of a finite longitudinal MR at $\theta =0$ rules out its semiclassical origin. (c) MR as a function of $B$ on a log scale for various temperatures. All the curves show quasilinear behavior of MR with applied field $B$ such that ${\rm MR} \propto B^m$ with $m \sim 1.13-1.16$. (d) Kohler's plot of transverse MR.
\label{fig4}}
\end{figure}

Magnetic field dependence of transverse MR ($B \perp I$) at different temperatures is shown in Fig.~\ref{fig4}(a). At low temperature ($T=4$K), CaCdSn shows an extremely large transverse MR of $7.44\times 10^3 \%$ without any sign of saturation even at high magnetic fields as large as $B=12$T. A similar large transverse MR has also been reported in isostructural topological semimetals RCdGe (R=Ca,Yb)\cite{CaCdGe_PRB_2017,YbCdGe_PRB_2019} as well as other topological semimetals \cite{MoAs2_PRB_2018,Cd3As2_PRL_2015,NdSb_PRB_2016,TlBiSSe_PRB_2015,YSb_SR_2016}. The transverse MR decreases significantly with increasing temperature, reaching a low value of $13\%$ at 100K and 12T [inset of Fig.~\ref{fig4}(a)]. Figure \ref{fig4}(b) shows the variation of MR with angle ($\theta$) between the current and magnetic field at $4$K. Interestingly, the MR does not vary much with $\theta$ and we obtain a positive, non-saturating and large longitudinal MR ($B||I$) of $1.71 \times 10^3\%$, similar to that reported in WTe$_2$\cite{WTe2_PRB_R_2015}. 

In order to determine the power-law dependence of MR $i.e.$ MR $\propto B^m$, we plot the field dependence of MR on a log scale in Fig.~\ref{fig4}(c). A quasi-linear behavior is observed with $m \sim 1.13-1.16$ at various temperatures. To understand whether classical methods account for such an MR, we consider well-known Kohler's rule, $i.e.$ MR = $\alpha (B/\rho_0)^m $, where $\alpha$ is a constant. The MR as a function of $B/\rho_0$ is shown in Fig.~\ref{fig4}(d) at various temperatures. Clearly, the MR at different temperatures is not merging into a single curve as required by the Kohler's rule, suggesting its violation in CaCdSn. There would be several reasons for such behavior including the existence of multiple pockets in the Fermi surface\cite{ZrSiS_PNAS_2017,WP_PRB_2017, Xu_2015,Multiband_Kohler}, differences in scattering rates of various carriers, among other possibilities\cite{McKenzie_Kohler}. 
 
 %% MR
 
 \section{Discussion and conclusion}\label{disc} 
 
We now discuss viable reasons for observing a large quasilinear MR in CaCdSn. A classical theory based on disorder-induced mobility fluctuation has been used to explain linear MR in several topological and non-topological materials, where the multiple scattering of high mobility carriers generate a drift velocity by low mobility islands perpendicular to the electronic cycloidal trajectories in an applied electric field \cite{Ag2Se_Nature_1997, Hu2008, Cd3As2_PRL_2015}. In this mechanism, the linear crossover magnetic field ($B_c$) and the amplitude of MR scale with mobility as $B_c \propto 1/\mu$ and MR $\propto \mu$, respectively \cite{Parish2003, Kozlova_2012}. However, the absence of linear crossover field and the high quality of our single crystals with large RRR value rules out this mechanism. Abrikosov's quantum theory can account for large linear MR in materials with linearly dispersive bands where the charge carriers are confined to the lowest Landau level\cite{Abrikosov_PRB_1999}. However, the quantum limits dictate that $n \ll (eB_c/\hbar c)^{3/2}$ and $T \ll (eB_c\hbar/m^*c)$, where $n$ and $m^*$ are the carrier density and effective mass, respectively\cite{Abrikosov_2000,Abrikosov_PRB_1998,Abrikosov_PRB_1999}. This yields a critical field $B_c$ of around 300T with $n=2.98 \times 10^{20}$ cm$^{-3}$. Since we observe quasilinear MR at very low magnetic field and find large Fermi pockets, the applicability of this method is also unlikely. A large nonsaturating linear MR has been recently observed in system with open Fermi surfaces \cite{PtBi2_arxiv_2020} and in charge/spin density wave materials with sharp edge Fermi surfaces \cite{Feng11201}. However, the absence of such Fermi surfaces in CaCdSn negates this scenario.

The large linear MR was recently reported in ultrahigh mobility samples of topological Dirac semimetal Cd$_3$As$_2$ where it was attributed to the magnetic field induced change in the Fermi surface\cite{Cd3AS2_nature_2015}. A similar MR has been observed in CaAgAs and YbCdGe which are isostructural to CaCdSn\cite{CaCdGe_PRB_2017, YbCdGe_PRB_2019}. Since CaCdSn hosts nontrivial band structure with trivial and nontrivial Fermi surface pockets and high mobility charge carriers similar to YbCdGe, the nontrivial band structure of CaCdSn may be responsible for the large quasilinear MR. However, to exactly pin down the mechanism responsible for large quasilinear MR, thickness-dependent transport studies might be helpful in a manner similar to the topological insulators \cite{Bi2Se3_TIs}. However, this is a demanding task requiring further carefully planned investigations.

In summary, we have explored the nature of the topological nodal-line semimetal state of CaCdSn via its magnetotransport properties measured on high-quality single crystals. The nodal band-crossings in CaCdSn lie above the Fermi level and yield a Fermi surface which consists of a large hole pocket and two small electron pockets. This unique Fermiology appears to drive interesting magnetotransport properties. A field induced metal-semiconductor-like crossover and a plateau in resistivity are observed at low temperatures. Transport in CaCdSn is dominated by hole carriers with high mobility. The large nonsaturating quasilinear MR we have observed in CaCdSn is similar to that seen in other topological semimetals, but its mechanism is not clear.

\section*{Acknowledgements} 

Research support from IIT Kanpur is gratefully acknowledged. ZH also acknowledges support from SERB India (Grant No. CRG/2018/000220). The work at Northeastern University is supported by the U.S. Department of Energy (DOE), Office of Science, Basic Energy Sciences Grant No. DE-FG02-07ER46352, and benefited from Northeastern University’s Advanced Scientific Computation Center and the National Energy Research Scientific Computing Center through DOE Grant No. DE-AC02-05CH11231. H. L. acknowledges Academia Sinica (Taiwan) for support under the Innovative Materials and Analysis Technology Exploration (AS-iMATE-107-11) program. A. A. acknowledges Science Education and Research Board (SERB) and Department of Science and Technology (DST) of the government of India for financial support.
 
\bibliography{ReferenceAll}

%merlin.mbs apsrev4-1.bst 2010-07-25 4.21a (PWD, AO, DPC) hacked
%Control: key (0)
%Control: author (8) initials jnrlst
%Control: editor formatted (1) identically to author
%Control: production of article title (-1) disabled
%Control: page (0) single
%Control: year (1) truncated
%Control: production of eprint (0) enabled
\begin{thebibliography}{71}%
\makeatletter
\providecommand \@ifxundefined [1]{%
 \@ifx{#1\undefined}
}%
\providecommand \@ifnum [1]{%
 \ifnum #1\expandafter \@firstoftwo
 \else \expandafter \@secondoftwo
 \fi
}%
\providecommand \@ifx [1]{%
 \ifx #1\expandafter \@firstoftwo
 \else \expandafter \@secondoftwo
 \fi
}%
\providecommand \natexlab [1]{#1}%
\providecommand \enquote  [1]{``#1''}%
\providecommand \bibnamefont  [1]{#1}%
\providecommand \bibfnamefont [1]{#1}%
\providecommand \citenamefont [1]{#1}%
\providecommand \href@noop [0]{\@secondoftwo}%
\providecommand \href [0]{\begingroup \@sanitize@url \@href}%
\providecommand \@href[1]{\@@startlink{#1}\@@href}%
\providecommand \@@href[1]{\endgroup#1\@@endlink}%
\providecommand \@sanitize@url [0]{\catcode `\\12\catcode `\$12\catcode
  `\&12\catcode `\#12\catcode `\^12\catcode `\_12\catcode `\%12\relax}%
\providecommand \@@startlink[1]{}%
\providecommand \@@endlink[0]{}%
\providecommand \url  [0]{\begingroup\@sanitize@url \@url }%
\providecommand \@url [1]{\endgroup\@href {#1}{\urlprefix }}%
\providecommand \urlprefix  [0]{URL }%
\providecommand \Eprint [0]{\href }%
\providecommand \doibase [0]{http://dx.doi.org/}%
\providecommand \selectlanguage [0]{\@gobble}%
\providecommand \bibinfo  [0]{\@secondoftwo}%
\providecommand \bibfield  [0]{\@secondoftwo}%
\providecommand \translation [1]{[#1]}%
\providecommand \BibitemOpen [0]{}%
\providecommand \bibitemStop [0]{}%
\providecommand \bibitemNoStop [0]{.\EOS\space}%
\providecommand \EOS [0]{\spacefactor3000\relax}%
\providecommand \BibitemShut  [1]{\csname bibitem#1\endcsname}%
\let\auto@bib@innerbib\@empty
%</preamble>
\bibitem [{\citenamefont {Bansil}\ \emph {et~al.}(2016)\citenamefont {Bansil},
  \citenamefont {Lin},\ and\ \citenamefont {Das}}]{bansil2016}%
  \BibitemOpen
  \bibfield  {author} {\bibinfo {author} {\bibfnamefont {A.}~\bibnamefont
  {Bansil}}, \bibinfo {author} {\bibfnamefont {H.}~\bibnamefont {Lin}}, \ and\
  \bibinfo {author} {\bibfnamefont {T.}~\bibnamefont {Das}},\ }\href {\doibase
  10.1103/RevModPhys.88.021004} {\bibfield  {journal} {\bibinfo  {journal}
  {Rev. Mod. Phys.}\ }\textbf {\bibinfo {volume} {88}},\ \bibinfo {pages}
  {021004} (\bibinfo {year} {2016})}\BibitemShut {NoStop}%
\bibitem [{\citenamefont {Hasan}\ and\ \citenamefont
  {Kane}(2010)}]{RevModPhys.82.3045}%
  \BibitemOpen
  \bibfield  {author} {\bibinfo {author} {\bibfnamefont {M.~Z.}\ \bibnamefont
  {Hasan}}\ and\ \bibinfo {author} {\bibfnamefont {C.~L.}\ \bibnamefont
  {Kane}},\ }\href {\doibase 10.1103/RevModPhys.82.3045} {\bibfield  {journal}
  {\bibinfo  {journal} {Rev. Mod. Phys.}\ }\textbf {\bibinfo {volume} {82}},\
  \bibinfo {pages} {3045} (\bibinfo {year} {2010})}\BibitemShut {NoStop}%
\bibitem [{\citenamefont {Armitage}\ \emph {et~al.}(2018)\citenamefont
  {Armitage}, \citenamefont {Mele},\ and\ \citenamefont
  {Vishwanath}}]{RevModPhys.90.015001}%
  \BibitemOpen
  \bibfield  {author} {\bibinfo {author} {\bibfnamefont {N.~P.}\ \bibnamefont
  {Armitage}}, \bibinfo {author} {\bibfnamefont {E.~J.}\ \bibnamefont {Mele}},
  \ and\ \bibinfo {author} {\bibfnamefont {A.}~\bibnamefont {Vishwanath}},\
  }\href {\doibase 10.1103/RevModPhys.90.015001} {\bibfield  {journal}
  {\bibinfo  {journal} {Rev. Mod. Phys.}\ }\textbf {\bibinfo {volume} {90}},\
  \bibinfo {pages} {015001} (\bibinfo {year} {2018})}\BibitemShut {NoStop}%
\bibitem [{\citenamefont {Zhang}\ \emph {et~al.}(2016)\citenamefont {Zhang},
  \citenamefont {Xu}, \citenamefont {Belopolski}, \citenamefont {Yuan},
  \citenamefont {Lin}, \citenamefont {Tong}, \citenamefont {Bian},
  \citenamefont {Alidoust}, \citenamefont {Lee}, \citenamefont {Huang},
  \citenamefont {Chang}, \citenamefont {Chang}, \citenamefont {Hsu},
  \citenamefont {Jeng}, \citenamefont {Neupane}, \citenamefont {Sanchez},
  \citenamefont {Zheng}, \citenamefont {Wang}, \citenamefont {Lin},
  \citenamefont {Zhang}, \citenamefont {Lu}, \citenamefont {Shen},
  \citenamefont {Neupert}, \citenamefont {Zahid~Hasan},\ and\ \citenamefont
  {Jia}}]{TaAs_Nature_comn_2016}%
  \BibitemOpen
  \bibfield  {author} {\bibinfo {author} {\bibfnamefont {C.-L.}\ \bibnamefont
  {Zhang}}, \bibinfo {author} {\bibfnamefont {S.-Y.}\ \bibnamefont {Xu}},
  \bibinfo {author} {\bibfnamefont {I.}~\bibnamefont {Belopolski}}, \bibinfo
  {author} {\bibfnamefont {Z.}~\bibnamefont {Yuan}}, \bibinfo {author}
  {\bibfnamefont {Z.}~\bibnamefont {Lin}}, \bibinfo {author} {\bibfnamefont
  {B.}~\bibnamefont {Tong}}, \bibinfo {author} {\bibfnamefont {G.}~\bibnamefont
  {Bian}}, \bibinfo {author} {\bibfnamefont {N.}~\bibnamefont {Alidoust}},
  \bibinfo {author} {\bibfnamefont {C.-C.}\ \bibnamefont {Lee}}, \bibinfo
  {author} {\bibfnamefont {S.-M.}\ \bibnamefont {Huang}}, \bibinfo {author}
  {\bibfnamefont {T.-R.}\ \bibnamefont {Chang}}, \bibinfo {author}
  {\bibfnamefont {G.}~\bibnamefont {Chang}}, \bibinfo {author} {\bibfnamefont
  {C.-H.}\ \bibnamefont {Hsu}}, \bibinfo {author} {\bibfnamefont {H.-T.}\
  \bibnamefont {Jeng}}, \bibinfo {author} {\bibfnamefont {M.}~\bibnamefont
  {Neupane}}, \bibinfo {author} {\bibfnamefont {D.~S.}\ \bibnamefont
  {Sanchez}}, \bibinfo {author} {\bibfnamefont {H.}~\bibnamefont {Zheng}},
  \bibinfo {author} {\bibfnamefont {J.}~\bibnamefont {Wang}}, \bibinfo {author}
  {\bibfnamefont {H.}~\bibnamefont {Lin}}, \bibinfo {author} {\bibfnamefont
  {C.}~\bibnamefont {Zhang}}, \bibinfo {author} {\bibfnamefont {H.-Z.}\
  \bibnamefont {Lu}}, \bibinfo {author} {\bibfnamefont {S.-Q.}\ \bibnamefont
  {Shen}}, \bibinfo {author} {\bibfnamefont {T.}~\bibnamefont {Neupert}},
  \bibinfo {author} {\bibfnamefont {M.}~\bibnamefont {Zahid~Hasan}}, \ and\
  \bibinfo {author} {\bibfnamefont {S.}~\bibnamefont {Jia}},\ }\href {\doibase
  10.1038/ncomms10735} {\bibfield  {journal} {\bibinfo  {journal} {Nat.
  Commun.}\ }\textbf {\bibinfo {volume} {7}},\ \bibinfo {pages} {10735}
  (\bibinfo {year} {2016})}\BibitemShut {NoStop}%
\bibitem [{\citenamefont {Xiong}\ \emph {et~al.}(2015)\citenamefont {Xiong},
  \citenamefont {Kushwaha}, \citenamefont {Liang}, \citenamefont {Krizan},
  \citenamefont {Hirschberger}, \citenamefont {Wang}, \citenamefont {Cava},\
  and\ \citenamefont {Ong}}]{Na3Bi_Science_2015}%
  \BibitemOpen
  \bibfield  {author} {\bibinfo {author} {\bibfnamefont {J.}~\bibnamefont
  {Xiong}}, \bibinfo {author} {\bibfnamefont {S.~K.}\ \bibnamefont {Kushwaha}},
  \bibinfo {author} {\bibfnamefont {T.}~\bibnamefont {Liang}}, \bibinfo
  {author} {\bibfnamefont {J.~W.}\ \bibnamefont {Krizan}}, \bibinfo {author}
  {\bibfnamefont {M.}~\bibnamefont {Hirschberger}}, \bibinfo {author}
  {\bibfnamefont {W.}~\bibnamefont {Wang}}, \bibinfo {author} {\bibfnamefont
  {R.~J.}\ \bibnamefont {Cava}}, \ and\ \bibinfo {author} {\bibfnamefont
  {N.~P.}\ \bibnamefont {Ong}},\ }\href {\doibase 10.1126/science.aac6089}
  {\bibfield  {journal} {\bibinfo  {journal} {Science}\ }\textbf {\bibinfo
  {volume} {350}},\ \bibinfo {pages} {413} (\bibinfo {year}
  {2015})}\BibitemShut {NoStop}%
\bibitem [{\citenamefont {Shekhar}\ \emph
  {et~al.}(2015{\natexlab{a}})\citenamefont {Shekhar}, \citenamefont {Nayak},
  \citenamefont {Sun}, \citenamefont {Schmidt}, \citenamefont {Nicklas},
  \citenamefont {Leermakers}, \citenamefont {Zeitler}, \citenamefont
  {Skourski}, \citenamefont {Wosnitza}, \citenamefont {Liu}, \citenamefont
  {Chen}, \citenamefont {Schnelle}, \citenamefont {Borrmann}, \citenamefont
  {Grin}, \citenamefont {Felser},\ and\ \citenamefont {Yan}}]{NbP}%
  \BibitemOpen
  \bibfield  {author} {\bibinfo {author} {\bibfnamefont {C.}~\bibnamefont
  {Shekhar}}, \bibinfo {author} {\bibfnamefont {A.~K.}\ \bibnamefont {Nayak}},
  \bibinfo {author} {\bibfnamefont {Y.}~\bibnamefont {Sun}}, \bibinfo {author}
  {\bibfnamefont {M.}~\bibnamefont {Schmidt}}, \bibinfo {author} {\bibfnamefont
  {M.}~\bibnamefont {Nicklas}}, \bibinfo {author} {\bibfnamefont
  {I.}~\bibnamefont {Leermakers}}, \bibinfo {author} {\bibfnamefont
  {U.}~\bibnamefont {Zeitler}}, \bibinfo {author} {\bibfnamefont
  {Y.}~\bibnamefont {Skourski}}, \bibinfo {author} {\bibfnamefont
  {J.}~\bibnamefont {Wosnitza}}, \bibinfo {author} {\bibfnamefont
  {Z.}~\bibnamefont {Liu}}, \bibinfo {author} {\bibfnamefont {Y.}~\bibnamefont
  {Chen}}, \bibinfo {author} {\bibfnamefont {W.}~\bibnamefont {Schnelle}},
  \bibinfo {author} {\bibfnamefont {H.}~\bibnamefont {Borrmann}}, \bibinfo
  {author} {\bibfnamefont {Y.}~\bibnamefont {Grin}}, \bibinfo {author}
  {\bibfnamefont {C.}~\bibnamefont {Felser}}, \ and\ \bibinfo {author}
  {\bibfnamefont {B.}~\bibnamefont {Yan}},\ }\href
  {http://dx.doi.org/10.1038/nphys3372} {\bibfield  {journal} {\bibinfo
  {journal} {Nat. Phys.}\ }\textbf {\bibinfo {volume} {11}},\ \bibinfo {pages}
  {645 EP } (\bibinfo {year} {2015}{\natexlab{a}})}\BibitemShut {NoStop}%
\bibitem [{\citenamefont {Liang}\ \emph {et~al.}(2018)\citenamefont {Liang},
  \citenamefont {Lin}, \citenamefont {Gibson}, \citenamefont {Kushwaha},
  \citenamefont {Liu}, \citenamefont {Wang}, \citenamefont {Xiong},
  \citenamefont {Sobota}, \citenamefont {Hashimoto}, \citenamefont {Kirchmann},
  \citenamefont {Shen}, \citenamefont {Cava},\ and\ \citenamefont
  {Ong}}]{ZrTe5_Nature_phy_2018}%
  \BibitemOpen
  \bibfield  {author} {\bibinfo {author} {\bibfnamefont {T.}~\bibnamefont
  {Liang}}, \bibinfo {author} {\bibfnamefont {J.}~\bibnamefont {Lin}}, \bibinfo
  {author} {\bibfnamefont {Q.}~\bibnamefont {Gibson}}, \bibinfo {author}
  {\bibfnamefont {S.}~\bibnamefont {Kushwaha}}, \bibinfo {author}
  {\bibfnamefont {M.}~\bibnamefont {Liu}}, \bibinfo {author} {\bibfnamefont
  {W.}~\bibnamefont {Wang}}, \bibinfo {author} {\bibfnamefont {H.}~\bibnamefont
  {Xiong}}, \bibinfo {author} {\bibfnamefont {J.~A.}\ \bibnamefont {Sobota}},
  \bibinfo {author} {\bibfnamefont {M.}~\bibnamefont {Hashimoto}}, \bibinfo
  {author} {\bibfnamefont {P.~S.}\ \bibnamefont {Kirchmann}}, \bibinfo {author}
  {\bibfnamefont {Z.-X.}\ \bibnamefont {Shen}}, \bibinfo {author}
  {\bibfnamefont {R.~J.}\ \bibnamefont {Cava}}, \ and\ \bibinfo {author}
  {\bibfnamefont {N.~P.}\ \bibnamefont {Ong}},\ }\href {\doibase
  10.1038/s41567-018-0078-z} {\bibfield  {journal} {\bibinfo  {journal} {Nat.
  Phys.}\ }\textbf {\bibinfo {volume} {14}},\ \bibinfo {pages} {451} (\bibinfo
  {year} {2018})}\BibitemShut {NoStop}%
\bibitem [{\citenamefont {Fang}\ \emph {et~al.}(2016)\citenamefont {Fang},
  \citenamefont {Weng}, \citenamefont {Dai},\ and\ \citenamefont
  {Fang}}]{fang2016}%
  \BibitemOpen
  \bibfield  {author} {\bibinfo {author} {\bibfnamefont {C.}~\bibnamefont
  {Fang}}, \bibinfo {author} {\bibfnamefont {H.}~\bibnamefont {Weng}}, \bibinfo
  {author} {\bibfnamefont {X.}~\bibnamefont {Dai}}, \ and\ \bibinfo {author}
  {\bibfnamefont {Z.}~\bibnamefont {Fang}},\ }\href {\doibase
  10.1088/1674-1056/25/11/117106} {\bibfield  {journal} {\bibinfo  {journal}
  {Chinese Physics B}\ }\textbf {\bibinfo {volume} {25}},\ \bibinfo {pages}
  {117106} (\bibinfo {year} {2016})}\BibitemShut {NoStop}%
\bibitem [{\citenamefont {Bian}\ \emph {et~al.}(2016)\citenamefont {Bian},
  \citenamefont {Chang}, \citenamefont {Sankar}, \citenamefont {Xu},
  \citenamefont {Zheng}, \citenamefont {Neupert}, \citenamefont {Chiu},
  \citenamefont {Huang}, \citenamefont {Chang}, \citenamefont {Belopolski},
  \citenamefont {Sanchez}, \citenamefont {Neupane}, \citenamefont {Alidoust},
  \citenamefont {Liu}, \citenamefont {Wang}, \citenamefont {Lee}, \citenamefont
  {Jeng}, \citenamefont {Zhang}, \citenamefont {Yuan}, \citenamefont {Jia},
  \citenamefont {Bansil}, \citenamefont {Chou}, \citenamefont {Lin},\ and\
  \citenamefont {Hasan}}]{bian2016}%
  \BibitemOpen
  \bibfield  {author} {\bibinfo {author} {\bibfnamefont {G.}~\bibnamefont
  {Bian}}, \bibinfo {author} {\bibfnamefont {T.-R.}\ \bibnamefont {Chang}},
  \bibinfo {author} {\bibfnamefont {R.}~\bibnamefont {Sankar}}, \bibinfo
  {author} {\bibfnamefont {S.-Y.}\ \bibnamefont {Xu}}, \bibinfo {author}
  {\bibfnamefont {H.}~\bibnamefont {Zheng}}, \bibinfo {author} {\bibfnamefont
  {T.}~\bibnamefont {Neupert}}, \bibinfo {author} {\bibfnamefont {C.-K.}\
  \bibnamefont {Chiu}}, \bibinfo {author} {\bibfnamefont {S.-M.}\ \bibnamefont
  {Huang}}, \bibinfo {author} {\bibfnamefont {G.}~\bibnamefont {Chang}},
  \bibinfo {author} {\bibfnamefont {I.}~\bibnamefont {Belopolski}}, \bibinfo
  {author} {\bibfnamefont {D.~S.}\ \bibnamefont {Sanchez}}, \bibinfo {author}
  {\bibfnamefont {M.}~\bibnamefont {Neupane}}, \bibinfo {author} {\bibfnamefont
  {N.}~\bibnamefont {Alidoust}}, \bibinfo {author} {\bibfnamefont
  {C.}~\bibnamefont {Liu}}, \bibinfo {author} {\bibfnamefont {B.}~\bibnamefont
  {Wang}}, \bibinfo {author} {\bibfnamefont {C.-C.}\ \bibnamefont {Lee}},
  \bibinfo {author} {\bibfnamefont {H.-T.}\ \bibnamefont {Jeng}}, \bibinfo
  {author} {\bibfnamefont {C.}~\bibnamefont {Zhang}}, \bibinfo {author}
  {\bibfnamefont {Z.}~\bibnamefont {Yuan}}, \bibinfo {author} {\bibfnamefont
  {S.}~\bibnamefont {Jia}}, \bibinfo {author} {\bibfnamefont {A.}~\bibnamefont
  {Bansil}}, \bibinfo {author} {\bibfnamefont {F.}~\bibnamefont {Chou}},
  \bibinfo {author} {\bibfnamefont {H.}~\bibnamefont {Lin}}, \ and\ \bibinfo
  {author} {\bibfnamefont {M.~Z.}\ \bibnamefont {Hasan}},\ }\href {\doibase
  10.1038/ncomms10556} {\bibfield  {journal} {\bibinfo  {journal} {Nat.
  Commun.}\ }\textbf {\bibinfo {volume} {7}},\ \bibinfo {pages} {10556}
  (\bibinfo {year} {2016})}\BibitemShut {NoStop}%
\bibitem [{\citenamefont {Wu}\ \emph {et~al.}(2016)\citenamefont {Wu},
  \citenamefont {Wang}, \citenamefont {Mun}, \citenamefont {Johnson},
  \citenamefont {Mou}, \citenamefont {Huang}, \citenamefont {Lee},
  \citenamefont {Bud'ko}, \citenamefont {Canfield},\ and\ \citenamefont
  {Kaminski}}]{PtSn4_natphy_2016}%
  \BibitemOpen
  \bibfield  {author} {\bibinfo {author} {\bibfnamefont {Y.}~\bibnamefont
  {Wu}}, \bibinfo {author} {\bibfnamefont {L.-L.}\ \bibnamefont {Wang}},
  \bibinfo {author} {\bibfnamefont {E.}~\bibnamefont {Mun}}, \bibinfo {author}
  {\bibfnamefont {D.~D.}\ \bibnamefont {Johnson}}, \bibinfo {author}
  {\bibfnamefont {D.}~\bibnamefont {Mou}}, \bibinfo {author} {\bibfnamefont
  {L.}~\bibnamefont {Huang}}, \bibinfo {author} {\bibfnamefont
  {Y.}~\bibnamefont {Lee}}, \bibinfo {author} {\bibfnamefont {S.~L.}\
  \bibnamefont {Bud'ko}}, \bibinfo {author} {\bibfnamefont {P.~C.}\
  \bibnamefont {Canfield}}, \ and\ \bibinfo {author} {\bibfnamefont
  {A.}~\bibnamefont {Kaminski}},\ }\href {https://doi.org/10.1038/nphys3712}
  {\bibfield  {journal} {\bibinfo  {journal} {Nat. Phys.}\ }\textbf {\bibinfo
  {volume} {12}},\ \bibinfo {pages} {667} (\bibinfo {year} {2016})}\BibitemShut
  {NoStop}%
\bibitem [{\citenamefont {Chang}\ \emph {et~al.}(2017)\citenamefont {Chang},
  \citenamefont {Xu}, \citenamefont {Zhou}, \citenamefont {Huang},
  \citenamefont {Singh}, \citenamefont {Wang}, \citenamefont {Belopolski},
  \citenamefont {Yin}, \citenamefont {Zhang}, \citenamefont {Bansil},
  \citenamefont {Lin},\ and\ \citenamefont {Hasan}}]{chang2017}%
  \BibitemOpen
  \bibfield  {author} {\bibinfo {author} {\bibfnamefont {G.}~\bibnamefont
  {Chang}}, \bibinfo {author} {\bibfnamefont {S.-Y.}\ \bibnamefont {Xu}},
  \bibinfo {author} {\bibfnamefont {X.}~\bibnamefont {Zhou}}, \bibinfo {author}
  {\bibfnamefont {S.-M.}\ \bibnamefont {Huang}}, \bibinfo {author}
  {\bibfnamefont {B.}~\bibnamefont {Singh}}, \bibinfo {author} {\bibfnamefont
  {B.}~\bibnamefont {Wang}}, \bibinfo {author} {\bibfnamefont {I.}~\bibnamefont
  {Belopolski}}, \bibinfo {author} {\bibfnamefont {J.}~\bibnamefont {Yin}},
  \bibinfo {author} {\bibfnamefont {S.}~\bibnamefont {Zhang}}, \bibinfo
  {author} {\bibfnamefont {A.}~\bibnamefont {Bansil}}, \bibinfo {author}
  {\bibfnamefont {H.}~\bibnamefont {Lin}}, \ and\ \bibinfo {author}
  {\bibfnamefont {M.~Z.}\ \bibnamefont {Hasan}},\ }\href {\doibase
  10.1103/PhysRevLett.119.156401} {\bibfield  {journal} {\bibinfo  {journal}
  {Phys. Rev. Lett.}\ }\textbf {\bibinfo {volume} {119}},\ \bibinfo {pages}
  {156401} (\bibinfo {year} {2017})}\BibitemShut {NoStop}%
\bibitem [{\citenamefont {Wang}\ \emph
  {et~al.}(2017{\natexlab{a}})\citenamefont {Wang}, \citenamefont {Ma},
  \citenamefont {Emmanouilidou}, \citenamefont {Shen}, \citenamefont {Hsu},
  \citenamefont {Zhou}, \citenamefont {Zuo}, \citenamefont {Song},
  \citenamefont {Xu}, \citenamefont {Wang}, \citenamefont {Huang},
  \citenamefont {Ni},\ and\ \citenamefont {Liu}}]{CaAgAs_PRB_R_2017}%
  \BibitemOpen
  \bibfield  {author} {\bibinfo {author} {\bibfnamefont {X.-B.}\ \bibnamefont
  {Wang}}, \bibinfo {author} {\bibfnamefont {X.-M.}\ \bibnamefont {Ma}},
  \bibinfo {author} {\bibfnamefont {E.}~\bibnamefont {Emmanouilidou}}, \bibinfo
  {author} {\bibfnamefont {B.}~\bibnamefont {Shen}}, \bibinfo {author}
  {\bibfnamefont {C.-H.}\ \bibnamefont {Hsu}}, \bibinfo {author} {\bibfnamefont
  {C.-S.}\ \bibnamefont {Zhou}}, \bibinfo {author} {\bibfnamefont
  {Y.}~\bibnamefont {Zuo}}, \bibinfo {author} {\bibfnamefont {R.-R.}\
  \bibnamefont {Song}}, \bibinfo {author} {\bibfnamefont {S.-Y.}\ \bibnamefont
  {Xu}}, \bibinfo {author} {\bibfnamefont {G.}~\bibnamefont {Wang}}, \bibinfo
  {author} {\bibfnamefont {L.}~\bibnamefont {Huang}}, \bibinfo {author}
  {\bibfnamefont {N.}~\bibnamefont {Ni}}, \ and\ \bibinfo {author}
  {\bibfnamefont {C.}~\bibnamefont {Liu}},\ }\href {\doibase
  10.1103/PhysRevB.96.161112} {\bibfield  {journal} {\bibinfo  {journal} {Phys.
  Rev. B}\ }\textbf {\bibinfo {volume} {96}},\ \bibinfo {pages} {161112}
  (\bibinfo {year} {2017}{\natexlab{a}})}\BibitemShut {NoStop}%
\bibitem [{\citenamefont {Singha}\ \emph
  {et~al.}(2017{\natexlab{a}})\citenamefont {Singha}, \citenamefont {Pariari},
  \citenamefont {Satpati},\ and\ \citenamefont {Mandal}}]{ZrSiS_PNAS_2017}%
  \BibitemOpen
  \bibfield  {author} {\bibinfo {author} {\bibfnamefont {R.}~\bibnamefont
  {Singha}}, \bibinfo {author} {\bibfnamefont {A.~K.}\ \bibnamefont {Pariari}},
  \bibinfo {author} {\bibfnamefont {B.}~\bibnamefont {Satpati}}, \ and\
  \bibinfo {author} {\bibfnamefont {P.}~\bibnamefont {Mandal}},\ }\href
  {\doibase 10.1073/pnas.1618004114} {\bibfield  {journal} {\bibinfo  {journal}
  {Proc. Natl. Acad. Sci.}\ } (\bibinfo {year} {2017}{\natexlab{a}}),\
  10.1073/pnas.1618004114}\BibitemShut {NoStop}%
\bibitem [{\citenamefont {Rudenko}\ \emph {et~al.}(2018)\citenamefont
  {Rudenko}, \citenamefont {Stepanov}, \citenamefont {Lichtenstein},\ and\
  \citenamefont {Katsnelson}}]{ZrSiS_PRL_2018}%
  \BibitemOpen
  \bibfield  {author} {\bibinfo {author} {\bibfnamefont {A.~N.}\ \bibnamefont
  {Rudenko}}, \bibinfo {author} {\bibfnamefont {E.~A.}\ \bibnamefont
  {Stepanov}}, \bibinfo {author} {\bibfnamefont {A.~I.}\ \bibnamefont
  {Lichtenstein}}, \ and\ \bibinfo {author} {\bibfnamefont {M.~I.}\
  \bibnamefont {Katsnelson}},\ }\href {\doibase 10.1103/PhysRevLett.120.216401}
  {\bibfield  {journal} {\bibinfo  {journal} {Phys. Rev. Lett.}\ }\textbf
  {\bibinfo {volume} {120}},\ \bibinfo {pages} {216401} (\bibinfo {year}
  {2018})}\BibitemShut {NoStop}%
\bibitem [{\citenamefont {Ali}\ \emph {et~al.}(2016)\citenamefont {Ali},
  \citenamefont {Schoop}, \citenamefont {Garg}, \citenamefont {Lippmann},
  \citenamefont {Lara}, \citenamefont {Lotsch},\ and\ \citenamefont
  {Parkin}}]{ZrSiS_ScienceAdv_2016}%
  \BibitemOpen
  \bibfield  {author} {\bibinfo {author} {\bibfnamefont {M.~N.}\ \bibnamefont
  {Ali}}, \bibinfo {author} {\bibfnamefont {L.~M.}\ \bibnamefont {Schoop}},
  \bibinfo {author} {\bibfnamefont {C.}~\bibnamefont {Garg}}, \bibinfo {author}
  {\bibfnamefont {J.~M.}\ \bibnamefont {Lippmann}}, \bibinfo {author}
  {\bibfnamefont {E.}~\bibnamefont {Lara}}, \bibinfo {author} {\bibfnamefont
  {B.}~\bibnamefont {Lotsch}}, \ and\ \bibinfo {author} {\bibfnamefont
  {S.~S.~P.}\ \bibnamefont {Parkin}},\ }\href {\doibase 10.1126/sciadv.1601742}
  {\bibfield  {journal} {\bibinfo  {journal} {Sci. Adv.}\ }\textbf {\bibinfo
  {volume} {2}} (\bibinfo {year} {2016}),\ 10.1126/sciadv.1601742}\BibitemShut
  {NoStop}%
\bibitem [{\citenamefont {Chang}\ \emph {et~al.}(2016)\citenamefont {Chang},
  \citenamefont {Chen}, \citenamefont {Bian}, \citenamefont {Huang},
  \citenamefont {Zheng}, \citenamefont {Neupert}, \citenamefont {Sankar},
  \citenamefont {Xu}, \citenamefont {Belopolski}, \citenamefont {Chang},
  \citenamefont {Wang}, \citenamefont {Chou}, \citenamefont {Bansil},
  \citenamefont {Jeng}, \citenamefont {Lin},\ and\ \citenamefont
  {Hasan}}]{PbTaSe2_PRB_2016}%
  \BibitemOpen
  \bibfield  {author} {\bibinfo {author} {\bibfnamefont {T.-R.}\ \bibnamefont
  {Chang}}, \bibinfo {author} {\bibfnamefont {P.-J.}\ \bibnamefont {Chen}},
  \bibinfo {author} {\bibfnamefont {G.}~\bibnamefont {Bian}}, \bibinfo {author}
  {\bibfnamefont {S.-M.}\ \bibnamefont {Huang}}, \bibinfo {author}
  {\bibfnamefont {H.}~\bibnamefont {Zheng}}, \bibinfo {author} {\bibfnamefont
  {T.}~\bibnamefont {Neupert}}, \bibinfo {author} {\bibfnamefont
  {R.}~\bibnamefont {Sankar}}, \bibinfo {author} {\bibfnamefont {S.-Y.}\
  \bibnamefont {Xu}}, \bibinfo {author} {\bibfnamefont {I.}~\bibnamefont
  {Belopolski}}, \bibinfo {author} {\bibfnamefont {G.}~\bibnamefont {Chang}},
  \bibinfo {author} {\bibfnamefont {B.}~\bibnamefont {Wang}}, \bibinfo {author}
  {\bibfnamefont {F.}~\bibnamefont {Chou}}, \bibinfo {author} {\bibfnamefont
  {A.}~\bibnamefont {Bansil}}, \bibinfo {author} {\bibfnamefont {H.-T.}\
  \bibnamefont {Jeng}}, \bibinfo {author} {\bibfnamefont {H.}~\bibnamefont
  {Lin}}, \ and\ \bibinfo {author} {\bibfnamefont {M.~Z.}\ \bibnamefont
  {Hasan}},\ }\href {\doibase 10.1103/PhysRevB.93.245130} {\bibfield  {journal}
  {\bibinfo  {journal} {Phys. Rev. B}\ }\textbf {\bibinfo {volume} {93}},\
  \bibinfo {pages} {245130} (\bibinfo {year} {2016})}\BibitemShut {NoStop}%
\bibitem [{\citenamefont {Neupane}\ \emph {et~al.}(2016)\citenamefont
  {Neupane}, \citenamefont {Belopolski}, \citenamefont {Hosen}, \citenamefont
  {Sanchez}, \citenamefont {Sankar}, \citenamefont {Szlawska}, \citenamefont
  {Xu}, \citenamefont {Dimitri}, \citenamefont {Dhakal}, \citenamefont
  {Maldonado}, \citenamefont {Oppeneer}, \citenamefont {Kaczorowski},
  \citenamefont {Chou}, \citenamefont {Hasan},\ and\ \citenamefont
  {Durakiewicz}}]{ZrSiS_PRB_R_2016}%
  \BibitemOpen
  \bibfield  {author} {\bibinfo {author} {\bibfnamefont {M.}~\bibnamefont
  {Neupane}}, \bibinfo {author} {\bibfnamefont {I.}~\bibnamefont {Belopolski}},
  \bibinfo {author} {\bibfnamefont {M.~M.}\ \bibnamefont {Hosen}}, \bibinfo
  {author} {\bibfnamefont {D.~S.}\ \bibnamefont {Sanchez}}, \bibinfo {author}
  {\bibfnamefont {R.}~\bibnamefont {Sankar}}, \bibinfo {author} {\bibfnamefont
  {M.}~\bibnamefont {Szlawska}}, \bibinfo {author} {\bibfnamefont {S.-Y.}\
  \bibnamefont {Xu}}, \bibinfo {author} {\bibfnamefont {K.}~\bibnamefont
  {Dimitri}}, \bibinfo {author} {\bibfnamefont {N.}~\bibnamefont {Dhakal}},
  \bibinfo {author} {\bibfnamefont {P.}~\bibnamefont {Maldonado}}, \bibinfo
  {author} {\bibfnamefont {P.~M.}\ \bibnamefont {Oppeneer}}, \bibinfo {author}
  {\bibfnamefont {D.}~\bibnamefont {Kaczorowski}}, \bibinfo {author}
  {\bibfnamefont {F.}~\bibnamefont {Chou}}, \bibinfo {author} {\bibfnamefont
  {M.~Z.}\ \bibnamefont {Hasan}}, \ and\ \bibinfo {author} {\bibfnamefont
  {T.}~\bibnamefont {Durakiewicz}},\ }\href {\doibase
  10.1103/PhysRevB.93.201104} {\bibfield  {journal} {\bibinfo  {journal} {Phys.
  Rev. B}\ }\textbf {\bibinfo {volume} {93}},\ \bibinfo {pages} {201104}
  (\bibinfo {year} {2016})}\BibitemShut {NoStop}%
\bibitem [{\citenamefont {Hosen}\ \emph {et~al.}(2017)\citenamefont {Hosen},
  \citenamefont {Dimitri}, \citenamefont {Belopolski}, \citenamefont
  {Maldonado}, \citenamefont {Sankar}, \citenamefont {Dhakal}, \citenamefont
  {Dhakal}, \citenamefont {Cole}, \citenamefont {Oppeneer}, \citenamefont
  {Kaczorowski}, \citenamefont {Chou}, \citenamefont {Hasan}, \citenamefont
  {Durakiewicz},\ and\ \citenamefont {Neupane}}]{ZrSiX_PRB_R_2017}%
  \BibitemOpen
  \bibfield  {author} {\bibinfo {author} {\bibfnamefont {M.~M.}\ \bibnamefont
  {Hosen}}, \bibinfo {author} {\bibfnamefont {K.}~\bibnamefont {Dimitri}},
  \bibinfo {author} {\bibfnamefont {I.}~\bibnamefont {Belopolski}}, \bibinfo
  {author} {\bibfnamefont {P.}~\bibnamefont {Maldonado}}, \bibinfo {author}
  {\bibfnamefont {R.}~\bibnamefont {Sankar}}, \bibinfo {author} {\bibfnamefont
  {N.}~\bibnamefont {Dhakal}}, \bibinfo {author} {\bibfnamefont
  {G.}~\bibnamefont {Dhakal}}, \bibinfo {author} {\bibfnamefont
  {T.}~\bibnamefont {Cole}}, \bibinfo {author} {\bibfnamefont {P.~M.}\
  \bibnamefont {Oppeneer}}, \bibinfo {author} {\bibfnamefont {D.}~\bibnamefont
  {Kaczorowski}}, \bibinfo {author} {\bibfnamefont {F.}~\bibnamefont {Chou}},
  \bibinfo {author} {\bibfnamefont {M.~Z.}\ \bibnamefont {Hasan}}, \bibinfo
  {author} {\bibfnamefont {T.}~\bibnamefont {Durakiewicz}}, \ and\ \bibinfo
  {author} {\bibfnamefont {M.}~\bibnamefont {Neupane}},\ }\href {\doibase
  10.1103/PhysRevB.95.161101} {\bibfield  {journal} {\bibinfo  {journal} {Phys.
  Rev. B}\ }\textbf {\bibinfo {volume} {95}},\ \bibinfo {pages} {161101}
  (\bibinfo {year} {2017})}\BibitemShut {NoStop}%
\bibitem [{\citenamefont {Singha}\ \emph {et~al.}(2018)\citenamefont {Singha},
  \citenamefont {Pariari}, \citenamefont {Gupta}, \citenamefont {Das},\ and\
  \citenamefont {Mandal}}]{MoAs2_PRB_2018}%
  \BibitemOpen
  \bibfield  {author} {\bibinfo {author} {\bibfnamefont {R.}~\bibnamefont
  {Singha}}, \bibinfo {author} {\bibfnamefont {A.}~\bibnamefont {Pariari}},
  \bibinfo {author} {\bibfnamefont {G.~K.}\ \bibnamefont {Gupta}}, \bibinfo
  {author} {\bibfnamefont {T.}~\bibnamefont {Das}}, \ and\ \bibinfo {author}
  {\bibfnamefont {P.}~\bibnamefont {Mandal}},\ }\href {\doibase
  10.1103/PhysRevB.97.155120} {\bibfield  {journal} {\bibinfo  {journal} {Phys.
  Rev. B}\ }\textbf {\bibinfo {volume} {97}},\ \bibinfo {pages} {155120}
  (\bibinfo {year} {2018})}\BibitemShut {NoStop}%
\bibitem [{\citenamefont {Narayanan}\ \emph {et~al.}(2015)\citenamefont
  {Narayanan}, \citenamefont {Watson}, \citenamefont {Blake}, \citenamefont
  {Bruyant}, \citenamefont {Drigo}, \citenamefont {Chen}, \citenamefont
  {Prabhakaran}, \citenamefont {Yan}, \citenamefont {Felser}, \citenamefont
  {Kong}, \citenamefont {Canfield},\ and\ \citenamefont
  {Coldea}}]{Cd3As2_PRL_2015}%
  \BibitemOpen
  \bibfield  {author} {\bibinfo {author} {\bibfnamefont {A.}~\bibnamefont
  {Narayanan}}, \bibinfo {author} {\bibfnamefont {M.~D.}\ \bibnamefont
  {Watson}}, \bibinfo {author} {\bibfnamefont {S.~F.}\ \bibnamefont {Blake}},
  \bibinfo {author} {\bibfnamefont {N.}~\bibnamefont {Bruyant}}, \bibinfo
  {author} {\bibfnamefont {L.}~\bibnamefont {Drigo}}, \bibinfo {author}
  {\bibfnamefont {Y.~L.}\ \bibnamefont {Chen}}, \bibinfo {author}
  {\bibfnamefont {D.}~\bibnamefont {Prabhakaran}}, \bibinfo {author}
  {\bibfnamefont {B.}~\bibnamefont {Yan}}, \bibinfo {author} {\bibfnamefont
  {C.}~\bibnamefont {Felser}}, \bibinfo {author} {\bibfnamefont
  {T.}~\bibnamefont {Kong}}, \bibinfo {author} {\bibfnamefont {P.~C.}\
  \bibnamefont {Canfield}}, \ and\ \bibinfo {author} {\bibfnamefont {A.~I.}\
  \bibnamefont {Coldea}},\ }\href {\doibase 10.1103/PhysRevLett.114.117201}
  {\bibfield  {journal} {\bibinfo  {journal} {Phys. Rev. Lett.}\ }\textbf
  {\bibinfo {volume} {114}},\ \bibinfo {pages} {117201} (\bibinfo {year}
  {2015})}\BibitemShut {NoStop}%
\bibitem [{\citenamefont {Wakeham}\ \emph {et~al.}(2016)\citenamefont
  {Wakeham}, \citenamefont {Bauer}, \citenamefont {Neupane},\ and\
  \citenamefont {Ronning}}]{NdSb_PRB_2016}%
  \BibitemOpen
  \bibfield  {author} {\bibinfo {author} {\bibfnamefont {N.}~\bibnamefont
  {Wakeham}}, \bibinfo {author} {\bibfnamefont {E.~D.}\ \bibnamefont {Bauer}},
  \bibinfo {author} {\bibfnamefont {M.}~\bibnamefont {Neupane}}, \ and\
  \bibinfo {author} {\bibfnamefont {F.}~\bibnamefont {Ronning}},\ }\href
  {\doibase 10.1103/PhysRevB.93.205152} {\bibfield  {journal} {\bibinfo
  {journal} {Phys. Rev. B}\ }\textbf {\bibinfo {volume} {93}},\ \bibinfo
  {pages} {205152} (\bibinfo {year} {2016})}\BibitemShut {NoStop}%
\bibitem [{\citenamefont {Novak}\ \emph {et~al.}(2015)\citenamefont {Novak},
  \citenamefont {Sasaki}, \citenamefont {Segawa},\ and\ \citenamefont
  {Ando}}]{TlBiSSe_PRB_2015}%
  \BibitemOpen
  \bibfield  {author} {\bibinfo {author} {\bibfnamefont {M.}~\bibnamefont
  {Novak}}, \bibinfo {author} {\bibfnamefont {S.}~\bibnamefont {Sasaki}},
  \bibinfo {author} {\bibfnamefont {K.}~\bibnamefont {Segawa}}, \ and\ \bibinfo
  {author} {\bibfnamefont {Y.}~\bibnamefont {Ando}},\ }\href {\doibase
  10.1103/PhysRevB.91.041203} {\bibfield  {journal} {\bibinfo  {journal} {Phys.
  Rev. B}\ }\textbf {\bibinfo {volume} {91}},\ \bibinfo {pages} {041203}
  (\bibinfo {year} {2015})}\BibitemShut {NoStop}%
\bibitem [{\citenamefont {Pavlosiuk}\ \emph {et~al.}(2016)\citenamefont
  {Pavlosiuk}, \citenamefont {Swatek},\ and\ \citenamefont
  {Wisniewski}}]{YSb_SR_2016}%
  \BibitemOpen
  \bibfield  {author} {\bibinfo {author} {\bibfnamefont {O.}~\bibnamefont
  {Pavlosiuk}}, \bibinfo {author} {\bibfnamefont {P.}~\bibnamefont {Swatek}}, \
  and\ \bibinfo {author} {\bibfnamefont {P.}~\bibnamefont {Wisniewski}},\
  }\href {https://doi.org/10.1038/srep38691} {\bibfield  {journal} {\bibinfo
  {journal} {Sci. Rep.}\ }\textbf {\bibinfo {volume} {6}},\ \bibinfo {pages}
  {38691} (\bibinfo {year} {2016})},\ \bibinfo {note} {article}\BibitemShut
  {NoStop}%
\bibitem [{\citenamefont {Emmanouilidou}\ \emph {et~al.}(2017)\citenamefont
  {Emmanouilidou}, \citenamefont {Shen}, \citenamefont {Deng}, \citenamefont
  {Chang}, \citenamefont {Shi}, \citenamefont {Kotliar}, \citenamefont {Xu},\
  and\ \citenamefont {Ni}}]{CaCdGe_PRB_2017}%
  \BibitemOpen
  \bibfield  {author} {\bibinfo {author} {\bibfnamefont {E.}~\bibnamefont
  {Emmanouilidou}}, \bibinfo {author} {\bibfnamefont {B.}~\bibnamefont {Shen}},
  \bibinfo {author} {\bibfnamefont {X.}~\bibnamefont {Deng}}, \bibinfo {author}
  {\bibfnamefont {T.-R.}\ \bibnamefont {Chang}}, \bibinfo {author}
  {\bibfnamefont {A.}~\bibnamefont {Shi}}, \bibinfo {author} {\bibfnamefont
  {G.}~\bibnamefont {Kotliar}}, \bibinfo {author} {\bibfnamefont {S.-Y.}\
  \bibnamefont {Xu}}, \ and\ \bibinfo {author} {\bibfnamefont {N.}~\bibnamefont
  {Ni}},\ }\href {\doibase 10.1103/PhysRevB.95.245113} {\bibfield  {journal}
  {\bibinfo  {journal} {Phys. Rev. B}\ }\textbf {\bibinfo {volume} {95}},\
  \bibinfo {pages} {245113} (\bibinfo {year} {2017})}\BibitemShut {NoStop}%
\bibitem [{\citenamefont {Laha}\ \emph {et~al.}(2019)\citenamefont {Laha},
  \citenamefont {Malick}, \citenamefont {Singha}, \citenamefont {Mandal},
  \citenamefont {Rambabu}, \citenamefont {Kanchana},\ and\ \citenamefont
  {Hossain}}]{YbCdGe_PRB_2019}%
  \BibitemOpen
  \bibfield  {author} {\bibinfo {author} {\bibfnamefont {A.}~\bibnamefont
  {Laha}}, \bibinfo {author} {\bibfnamefont {S.}~\bibnamefont {Malick}},
  \bibinfo {author} {\bibfnamefont {R.}~\bibnamefont {Singha}}, \bibinfo
  {author} {\bibfnamefont {P.}~\bibnamefont {Mandal}}, \bibinfo {author}
  {\bibfnamefont {P.}~\bibnamefont {Rambabu}}, \bibinfo {author} {\bibfnamefont
  {V.}~\bibnamefont {Kanchana}}, \ and\ \bibinfo {author} {\bibfnamefont
  {Z.}~\bibnamefont {Hossain}},\ }\href {\doibase 10.1103/PhysRevB.99.241102}
  {\bibfield  {journal} {\bibinfo  {journal} {Phys. Rev. B}\ }\textbf {\bibinfo
  {volume} {99}},\ \bibinfo {pages} {241102} (\bibinfo {year}
  {2019})}\BibitemShut {NoStop}%
\bibitem [{\citenamefont {Hu}\ and\ \citenamefont {Rosenbaum}(2008)}]{Hu2008}%
  \BibitemOpen
  \bibfield  {author} {\bibinfo {author} {\bibfnamefont {J.}~\bibnamefont
  {Hu}}\ and\ \bibinfo {author} {\bibfnamefont {T.~F.}\ \bibnamefont
  {Rosenbaum}},\ }\href {https://doi.org/10.1038/nmat2259} {\bibfield
  {journal} {\bibinfo  {journal} {Nature Materials}\ }\textbf {\bibinfo
  {volume} {7}},\ \bibinfo {pages} {697} (\bibinfo {year} {2008})}\BibitemShut
  {NoStop}%
\bibitem [{\citenamefont {Leahy}\ \emph {et~al.}(2018)\citenamefont {Leahy},
  \citenamefont {Lin}, \citenamefont {Siegfried}, \citenamefont {Treglia},
  \citenamefont {Song}, \citenamefont {Nandkishore},\ and\ \citenamefont
  {Lee}}]{Leahy10570}%
  \BibitemOpen
  \bibfield  {author} {\bibinfo {author} {\bibfnamefont {I.~A.}\ \bibnamefont
  {Leahy}}, \bibinfo {author} {\bibfnamefont {Y.-P.}\ \bibnamefont {Lin}},
  \bibinfo {author} {\bibfnamefont {P.~E.}\ \bibnamefont {Siegfried}}, \bibinfo
  {author} {\bibfnamefont {A.~C.}\ \bibnamefont {Treglia}}, \bibinfo {author}
  {\bibfnamefont {J.~C.~W.}\ \bibnamefont {Song}}, \bibinfo {author}
  {\bibfnamefont {R.~M.}\ \bibnamefont {Nandkishore}}, \ and\ \bibinfo {author}
  {\bibfnamefont {M.}~\bibnamefont {Lee}},\ }\href {\doibase
  10.1073/pnas.1808747115} {\bibfield  {journal} {\bibinfo  {journal} {Proc.
  Natl. Acad. Sci.}\ }\textbf {\bibinfo {volume} {115}},\ \bibinfo {pages}
  {10570} (\bibinfo {year} {2018})}\BibitemShut {NoStop}%
\bibitem [{\citenamefont {Shekhar}\ \emph
  {et~al.}(2015{\natexlab{b}})\citenamefont {Shekhar}, \citenamefont {Nayak},
  \citenamefont {Sun}, \citenamefont {Schmidt}, \citenamefont {Nicklas},
  \citenamefont {Leermakers}, \citenamefont {Zeitler}, \citenamefont
  {Skourski}, \citenamefont {Wosnitza}, \citenamefont {Liu}, \citenamefont
  {Chen}, \citenamefont {Schnelle}, \citenamefont {Borrmann}, \citenamefont
  {Grin}, \citenamefont {Felser},\ and\ \citenamefont {Yan}}]{NbP_natphy_2015}%
  \BibitemOpen
  \bibfield  {author} {\bibinfo {author} {\bibfnamefont {C.}~\bibnamefont
  {Shekhar}}, \bibinfo {author} {\bibfnamefont {A.~K.}\ \bibnamefont {Nayak}},
  \bibinfo {author} {\bibfnamefont {Y.}~\bibnamefont {Sun}}, \bibinfo {author}
  {\bibfnamefont {M.}~\bibnamefont {Schmidt}}, \bibinfo {author} {\bibfnamefont
  {M.}~\bibnamefont {Nicklas}}, \bibinfo {author} {\bibfnamefont
  {I.}~\bibnamefont {Leermakers}}, \bibinfo {author} {\bibfnamefont
  {U.}~\bibnamefont {Zeitler}}, \bibinfo {author} {\bibfnamefont
  {Y.}~\bibnamefont {Skourski}}, \bibinfo {author} {\bibfnamefont
  {J.}~\bibnamefont {Wosnitza}}, \bibinfo {author} {\bibfnamefont
  {Z.}~\bibnamefont {Liu}}, \bibinfo {author} {\bibfnamefont {Y.}~\bibnamefont
  {Chen}}, \bibinfo {author} {\bibfnamefont {W.}~\bibnamefont {Schnelle}},
  \bibinfo {author} {\bibfnamefont {H.}~\bibnamefont {Borrmann}}, \bibinfo
  {author} {\bibfnamefont {Y.}~\bibnamefont {Grin}}, \bibinfo {author}
  {\bibfnamefont {C.}~\bibnamefont {Felser}}, \ and\ \bibinfo {author}
  {\bibfnamefont {B.}~\bibnamefont {Yan}},\ }\href {\doibase 10.1038/nphys3372}
  {\bibfield  {journal} {\bibinfo  {journal} {Nature Physics}\ }\textbf
  {\bibinfo {volume} {11}},\ \bibinfo {pages} {645} (\bibinfo {year}
  {2015}{\natexlab{b}})}\BibitemShut {NoStop}%
\bibitem [{\citenamefont {Gao}\ \emph {et~al.}(2017)\citenamefont {Gao},
  \citenamefont {Hao}, \citenamefont {Zheng}, \citenamefont {Ning},
  \citenamefont {Wu}, \citenamefont {Zhu}, \citenamefont {Zheng}, \citenamefont
  {Zhang}, \citenamefont {Lu}, \citenamefont {Zhang}, \citenamefont {Xi},
  \citenamefont {Yang}, \citenamefont {Du}, \citenamefont {Zhang},
  \citenamefont {Zhang},\ and\ \citenamefont {Tian}}]{PtBi2_PRL_2017}%
  \BibitemOpen
  \bibfield  {author} {\bibinfo {author} {\bibfnamefont {W.}~\bibnamefont
  {Gao}}, \bibinfo {author} {\bibfnamefont {N.}~\bibnamefont {Hao}}, \bibinfo
  {author} {\bibfnamefont {F.-W.}\ \bibnamefont {Zheng}}, \bibinfo {author}
  {\bibfnamefont {W.}~\bibnamefont {Ning}}, \bibinfo {author} {\bibfnamefont
  {M.}~\bibnamefont {Wu}}, \bibinfo {author} {\bibfnamefont {X.}~\bibnamefont
  {Zhu}}, \bibinfo {author} {\bibfnamefont {G.}~\bibnamefont {Zheng}}, \bibinfo
  {author} {\bibfnamefont {J.}~\bibnamefont {Zhang}}, \bibinfo {author}
  {\bibfnamefont {J.}~\bibnamefont {Lu}}, \bibinfo {author} {\bibfnamefont
  {H.}~\bibnamefont {Zhang}}, \bibinfo {author} {\bibfnamefont
  {C.}~\bibnamefont {Xi}}, \bibinfo {author} {\bibfnamefont {J.}~\bibnamefont
  {Yang}}, \bibinfo {author} {\bibfnamefont {H.}~\bibnamefont {Du}}, \bibinfo
  {author} {\bibfnamefont {P.}~\bibnamefont {Zhang}}, \bibinfo {author}
  {\bibfnamefont {Y.}~\bibnamefont {Zhang}}, \ and\ \bibinfo {author}
  {\bibfnamefont {M.}~\bibnamefont {Tian}},\ }\href {\doibase
  10.1103/PhysRevLett.118.256601} {\bibfield  {journal} {\bibinfo  {journal}
  {Phys. Rev. Lett.}\ }\textbf {\bibinfo {volume} {118}},\ \bibinfo {pages}
  {256601} (\bibinfo {year} {2017})}\BibitemShut {NoStop}%
\bibitem [{\citenamefont {Luo}\ \emph {et~al.}(2018)\citenamefont {Luo},
  \citenamefont {Xiao}, \citenamefont {Chen}, \citenamefont {Yan},
  \citenamefont {Pei}, \citenamefont {Sun}, \citenamefont {Lu}, \citenamefont
  {Tong}, \citenamefont {Sheng}, \citenamefont {Zhu}, \citenamefont {Song},\
  and\ \citenamefont {Sun}}]{PtSn4_PRB_2018}%
  \BibitemOpen
  \bibfield  {author} {\bibinfo {author} {\bibfnamefont {X.}~\bibnamefont
  {Luo}}, \bibinfo {author} {\bibfnamefont {R.~C.}\ \bibnamefont {Xiao}},
  \bibinfo {author} {\bibfnamefont {F.~C.}\ \bibnamefont {Chen}}, \bibinfo
  {author} {\bibfnamefont {J.}~\bibnamefont {Yan}}, \bibinfo {author}
  {\bibfnamefont {Q.~L.}\ \bibnamefont {Pei}}, \bibinfo {author} {\bibfnamefont
  {Y.}~\bibnamefont {Sun}}, \bibinfo {author} {\bibfnamefont {W.~J.}\
  \bibnamefont {Lu}}, \bibinfo {author} {\bibfnamefont {P.}~\bibnamefont
  {Tong}}, \bibinfo {author} {\bibfnamefont {Z.~G.}\ \bibnamefont {Sheng}},
  \bibinfo {author} {\bibfnamefont {X.~B.}\ \bibnamefont {Zhu}}, \bibinfo
  {author} {\bibfnamefont {W.~H.}\ \bibnamefont {Song}}, \ and\ \bibinfo
  {author} {\bibfnamefont {Y.~P.}\ \bibnamefont {Sun}},\ }\href {\doibase
  10.1103/PhysRevB.97.205132} {\bibfield  {journal} {\bibinfo  {journal} {Phys.
  Rev. B}\ }\textbf {\bibinfo {volume} {97}},\ \bibinfo {pages} {205132}
  (\bibinfo {year} {2018})}\BibitemShut {NoStop}%
\bibitem [{\citenamefont {Wang}\ \emph {et~al.}(2016)\citenamefont {Wang},
  \citenamefont {Yu}, \citenamefont {Guo}, \citenamefont {Liu},\ and\
  \citenamefont {Xia}}]{NbAs2TaAs2_PRB(R)_2016}%
  \BibitemOpen
  \bibfield  {author} {\bibinfo {author} {\bibfnamefont {Y.-Y.}\ \bibnamefont
  {Wang}}, \bibinfo {author} {\bibfnamefont {Q.-H.}\ \bibnamefont {Yu}},
  \bibinfo {author} {\bibfnamefont {P.-J.}\ \bibnamefont {Guo}}, \bibinfo
  {author} {\bibfnamefont {K.}~\bibnamefont {Liu}}, \ and\ \bibinfo {author}
  {\bibfnamefont {T.-L.}\ \bibnamefont {Xia}},\ }\href {\doibase
  10.1103/PhysRevB.94.041103} {\bibfield  {journal} {\bibinfo  {journal} {Phys.
  Rev. B}\ }\textbf {\bibinfo {volume} {94}},\ \bibinfo {pages} {041103}
  (\bibinfo {year} {2016})}\BibitemShut {NoStop}%
\bibitem [{\citenamefont {Yuan}\ \emph {et~al.}(2016)\citenamefont {Yuan},
  \citenamefont {Lu}, \citenamefont {Liu}, \citenamefont {Wang},\ and\
  \citenamefont {Jia}}]{NbAs2TaAs2_PRB_2016}%
  \BibitemOpen
  \bibfield  {author} {\bibinfo {author} {\bibfnamefont {Z.}~\bibnamefont
  {Yuan}}, \bibinfo {author} {\bibfnamefont {H.}~\bibnamefont {Lu}}, \bibinfo
  {author} {\bibfnamefont {Y.}~\bibnamefont {Liu}}, \bibinfo {author}
  {\bibfnamefont {J.}~\bibnamefont {Wang}}, \ and\ \bibinfo {author}
  {\bibfnamefont {S.}~\bibnamefont {Jia}},\ }\href {\doibase
  10.1103/PhysRevB.93.184405} {\bibfield  {journal} {\bibinfo  {journal} {Phys.
  Rev. B}\ }\textbf {\bibinfo {volume} {93}},\ \bibinfo {pages} {184405}
  (\bibinfo {year} {2016})}\BibitemShut {NoStop}%
\bibitem [{\citenamefont {Liang}\ \emph {et~al.}(2015)\citenamefont {Liang},
  \citenamefont {Gibson}, \citenamefont {Ali}, \citenamefont {Liu},
  \citenamefont {Cava},\ and\ \citenamefont {Ong}}]{Cd3As2_natmat_2015}%
  \BibitemOpen
  \bibfield  {author} {\bibinfo {author} {\bibfnamefont {T.}~\bibnamefont
  {Liang}}, \bibinfo {author} {\bibfnamefont {Q.}~\bibnamefont {Gibson}},
  \bibinfo {author} {\bibfnamefont {M.~N.}\ \bibnamefont {Ali}}, \bibinfo
  {author} {\bibfnamefont {M.}~\bibnamefont {Liu}}, \bibinfo {author}
  {\bibfnamefont {R.~J.}\ \bibnamefont {Cava}}, \ and\ \bibinfo {author}
  {\bibfnamefont {N.~P.}\ \bibnamefont {Ong}},\ }\href {\doibase
  10.1038/nmat4143} {\bibfield  {journal} {\bibinfo  {journal} {Nature
  Materials}\ }\textbf {\bibinfo {volume} {14}},\ \bibinfo {pages} {280}
  (\bibinfo {year} {2015})}\BibitemShut {NoStop}%
\bibitem [{\citenamefont {Ali}\ \emph {et~al.}(2014)\citenamefont {Ali},
  \citenamefont {Xiong}, \citenamefont {Flynn}, \citenamefont {Tao},
  \citenamefont {Gibson}, \citenamefont {Schoop}, \citenamefont {Liang},
  \citenamefont {Haldolaarachchige}, \citenamefont {Hirschberger},
  \citenamefont {Ong},\ and\ \citenamefont {Cava}}]{WTe2}%
  \BibitemOpen
  \bibfield  {author} {\bibinfo {author} {\bibfnamefont {M.~N.}\ \bibnamefont
  {Ali}}, \bibinfo {author} {\bibfnamefont {J.}~\bibnamefont {Xiong}}, \bibinfo
  {author} {\bibfnamefont {S.}~\bibnamefont {Flynn}}, \bibinfo {author}
  {\bibfnamefont {J.}~\bibnamefont {Tao}}, \bibinfo {author} {\bibfnamefont
  {Q.~D.}\ \bibnamefont {Gibson}}, \bibinfo {author} {\bibfnamefont {L.~M.}\
  \bibnamefont {Schoop}}, \bibinfo {author} {\bibfnamefont {T.}~\bibnamefont
  {Liang}}, \bibinfo {author} {\bibfnamefont {N.}~\bibnamefont
  {Haldolaarachchige}}, \bibinfo {author} {\bibfnamefont {M.}~\bibnamefont
  {Hirschberger}}, \bibinfo {author} {\bibfnamefont {N.~P.}\ \bibnamefont
  {Ong}}, \ and\ \bibinfo {author} {\bibfnamefont {R.~J.}\ \bibnamefont
  {Cava}},\ }\href {http://dx.doi.org/10.1038/nature13763} {\bibfield
  {journal} {\bibinfo  {journal} {Nature}\ }\textbf {\bibinfo {volume} {514}},\
  \bibinfo {pages} {205 EP } (\bibinfo {year} {2014})}\BibitemShut {NoStop}%
\bibitem [{\citenamefont {Kumar}\ \emph {et~al.}(2017)\citenamefont {Kumar},
  \citenamefont {Sun}, \citenamefont {Xu}, \citenamefont {Manna}, \citenamefont
  {Yao}, \citenamefont {S{\"u}ss}, \citenamefont {Leermakers}, \citenamefont
  {Young}, \citenamefont {F{\"o}rster}, \citenamefont {Schmidt}, \citenamefont
  {Borrmann}, \citenamefont {Yan}, \citenamefont {Zeitler}, \citenamefont
  {Shi}, \citenamefont {Felser},\ and\ \citenamefont
  {Shekhar}}]{WP2_MoP2_natcom_2017}%
  \BibitemOpen
  \bibfield  {author} {\bibinfo {author} {\bibfnamefont {N.}~\bibnamefont
  {Kumar}}, \bibinfo {author} {\bibfnamefont {Y.}~\bibnamefont {Sun}}, \bibinfo
  {author} {\bibfnamefont {N.}~\bibnamefont {Xu}}, \bibinfo {author}
  {\bibfnamefont {K.}~\bibnamefont {Manna}}, \bibinfo {author} {\bibfnamefont
  {M.}~\bibnamefont {Yao}}, \bibinfo {author} {\bibfnamefont {V.}~\bibnamefont
  {S{\"u}ss}}, \bibinfo {author} {\bibfnamefont {I.}~\bibnamefont
  {Leermakers}}, \bibinfo {author} {\bibfnamefont {O.}~\bibnamefont {Young}},
  \bibinfo {author} {\bibfnamefont {T.}~\bibnamefont {F{\"o}rster}}, \bibinfo
  {author} {\bibfnamefont {M.}~\bibnamefont {Schmidt}}, \bibinfo {author}
  {\bibfnamefont {H.}~\bibnamefont {Borrmann}}, \bibinfo {author}
  {\bibfnamefont {B.}~\bibnamefont {Yan}}, \bibinfo {author} {\bibfnamefont
  {U.}~\bibnamefont {Zeitler}}, \bibinfo {author} {\bibfnamefont
  {M.}~\bibnamefont {Shi}}, \bibinfo {author} {\bibfnamefont {C.}~\bibnamefont
  {Felser}}, \ and\ \bibinfo {author} {\bibfnamefont {C.}~\bibnamefont
  {Shekhar}},\ }\href {\doibase 10.1038/s41467-017-01758-z} {\bibfield
  {journal} {\bibinfo  {journal} {Nature Communications}\ }\textbf {\bibinfo
  {volume} {8}},\ \bibinfo {pages} {1642} (\bibinfo {year} {2017})}\BibitemShut
  {NoStop}%
\bibitem [{\citenamefont {Spain}\ and\ \citenamefont
  {Dillon}(1976)}]{SPAIN197623}%
  \BibitemOpen
  \bibfield  {author} {\bibinfo {author} {\bibfnamefont {I.}~\bibnamefont
  {Spain}}\ and\ \bibinfo {author} {\bibfnamefont {R.}~\bibnamefont {Dillon}},\
  }\href {\doibase https://doi.org/10.1016/0008-6223(76)90077-4} {\bibfield
  {journal} {\bibinfo  {journal} {Carbon}\ }\textbf {\bibinfo {volume} {14}},\
  \bibinfo {pages} {23 } (\bibinfo {year} {1976})}\BibitemShut {NoStop}%
\bibitem [{\citenamefont {Wang}\ \emph {et~al.}(2015)\citenamefont {Wang},
  \citenamefont {Thoutam}, \citenamefont {Xiao}, \citenamefont {Hu},
  \citenamefont {Das}, \citenamefont {Mao}, \citenamefont {Wei}, \citenamefont
  {Divan}, \citenamefont {Luican-Mayer}, \citenamefont {Crabtree},\ and\
  \citenamefont {Kwok}}]{WTe2_PRB_2015}%
  \BibitemOpen
  \bibfield  {author} {\bibinfo {author} {\bibfnamefont {Y.~L.}\ \bibnamefont
  {Wang}}, \bibinfo {author} {\bibfnamefont {L.~R.}\ \bibnamefont {Thoutam}},
  \bibinfo {author} {\bibfnamefont {Z.~L.}\ \bibnamefont {Xiao}}, \bibinfo
  {author} {\bibfnamefont {J.}~\bibnamefont {Hu}}, \bibinfo {author}
  {\bibfnamefont {S.}~\bibnamefont {Das}}, \bibinfo {author} {\bibfnamefont
  {Z.~Q.}\ \bibnamefont {Mao}}, \bibinfo {author} {\bibfnamefont
  {J.}~\bibnamefont {Wei}}, \bibinfo {author} {\bibfnamefont {R.}~\bibnamefont
  {Divan}}, \bibinfo {author} {\bibfnamefont {A.}~\bibnamefont {Luican-Mayer}},
  \bibinfo {author} {\bibfnamefont {G.~W.}\ \bibnamefont {Crabtree}}, \ and\
  \bibinfo {author} {\bibfnamefont {W.~K.}\ \bibnamefont {Kwok}},\ }\href
  {\doibase 10.1103/PhysRevB.92.180402} {\bibfield  {journal} {\bibinfo
  {journal} {Phys. Rev. B}\ }\textbf {\bibinfo {volume} {92}},\ \bibinfo
  {pages} {180402} (\bibinfo {year} {2015})}\BibitemShut {NoStop}%
\bibitem [{\citenamefont {Parish}\ and\ \citenamefont
  {Littlewood}(2003)}]{Parish2003}%
  \BibitemOpen
  \bibfield  {author} {\bibinfo {author} {\bibfnamefont {M.~M.}\ \bibnamefont
  {Parish}}\ and\ \bibinfo {author} {\bibfnamefont {P.~B.}\ \bibnamefont
  {Littlewood}},\ }\href {\doibase 10.1038/nature02073} {\bibfield  {journal}
  {\bibinfo  {journal} {Nature}\ }\textbf {\bibinfo {volume} {426}},\ \bibinfo
  {pages} {162} (\bibinfo {year} {2003})}\BibitemShut {NoStop}%
\bibitem [{\citenamefont {Song}\ \emph {et~al.}(2015)\citenamefont {Song},
  \citenamefont {Refael},\ and\ \citenamefont {Lee}}]{PhysRevB.92.180204}%
  \BibitemOpen
  \bibfield  {author} {\bibinfo {author} {\bibfnamefont {J.~C.~W.}\
  \bibnamefont {Song}}, \bibinfo {author} {\bibfnamefont {G.}~\bibnamefont
  {Refael}}, \ and\ \bibinfo {author} {\bibfnamefont {P.~A.}\ \bibnamefont
  {Lee}},\ }\href {\doibase 10.1103/PhysRevB.92.180204} {\bibfield  {journal}
  {\bibinfo  {journal} {Phys. Rev. B}\ }\textbf {\bibinfo {volume} {92}},\
  \bibinfo {pages} {180204} (\bibinfo {year} {2015})}\BibitemShut {NoStop}%
\bibitem [{\citenamefont {Abrikosov}(2000)}]{Abrikosov_2000}%
  \BibitemOpen
  \bibfield  {author} {\bibinfo {author} {\bibfnamefont {A.~A.}\ \bibnamefont
  {Abrikosov}},\ }\href {\doibase 10.1209/epl/i2000-00220-2} {\bibfield
  {journal} {\bibinfo  {journal} {Europhysics Letters ({EPL})}\ }\textbf
  {\bibinfo {volume} {49}},\ \bibinfo {pages} {789} (\bibinfo {year}
  {2000})}\BibitemShut {NoStop}%
\bibitem [{\citenamefont {Abrikosov}(1998)}]{Abrikosov_PRB_1998}%
  \BibitemOpen
  \bibfield  {author} {\bibinfo {author} {\bibfnamefont {A.~A.}\ \bibnamefont
  {Abrikosov}},\ }\href {\doibase 10.1103/PhysRevB.58.2788} {\bibfield
  {journal} {\bibinfo  {journal} {Phys. Rev. B}\ }\textbf {\bibinfo {volume}
  {58}},\ \bibinfo {pages} {2788} (\bibinfo {year} {1998})}\BibitemShut
  {NoStop}%
\bibitem [{\citenamefont {Abrikosov}(1999)}]{Abrikosov_PRB_1999}%
  \BibitemOpen
  \bibfield  {author} {\bibinfo {author} {\bibfnamefont {A.~A.}\ \bibnamefont
  {Abrikosov}},\ }\href {\doibase 10.1103/PhysRevB.60.4231} {\bibfield
  {journal} {\bibinfo  {journal} {Phys. Rev. B}\ }\textbf {\bibinfo {volume}
  {60}},\ \bibinfo {pages} {4231} (\bibinfo {year} {1999})}\BibitemShut
  {NoStop}%
\bibitem [{\citenamefont {Niu}\ \emph {et~al.}(2017)\citenamefont {Niu},
  \citenamefont {Yu}, \citenamefont {Yip}, \citenamefont {Lim}, \citenamefont
  {Kotegawa}, \citenamefont {Matsuoka}, \citenamefont {Sugawara}, \citenamefont
  {Tou}, \citenamefont {Yanase},\ and\ \citenamefont
  {Goh}}]{CrAs_Nature_Comn_2017}%
  \BibitemOpen
  \bibfield  {author} {\bibinfo {author} {\bibfnamefont {Q.}~\bibnamefont
  {Niu}}, \bibinfo {author} {\bibfnamefont {W.~C.}\ \bibnamefont {Yu}},
  \bibinfo {author} {\bibfnamefont {K.~Y.}\ \bibnamefont {Yip}}, \bibinfo
  {author} {\bibfnamefont {Z.~L.}\ \bibnamefont {Lim}}, \bibinfo {author}
  {\bibfnamefont {H.}~\bibnamefont {Kotegawa}}, \bibinfo {author}
  {\bibfnamefont {E.}~\bibnamefont {Matsuoka}}, \bibinfo {author}
  {\bibfnamefont {H.}~\bibnamefont {Sugawara}}, \bibinfo {author}
  {\bibfnamefont {H.}~\bibnamefont {Tou}}, \bibinfo {author} {\bibfnamefont
  {Y.}~\bibnamefont {Yanase}}, \ and\ \bibinfo {author} {\bibfnamefont {S.~K.}\
  \bibnamefont {Goh}},\ }\href {\doibase 10.1038/ncomms15358} {\bibfield
  {journal} {\bibinfo  {journal} {Nat. Commun.}\ }\textbf {\bibinfo {volume}
  {8}},\ \bibinfo {pages} {15358} (\bibinfo {year} {2017})}\BibitemShut
  {NoStop}%
\bibitem [{\citenamefont {Rodr{\'i}guez-Carvajal}(1993)}]{Fullprof_1993}%
  \BibitemOpen
  \bibfield  {author} {\bibinfo {author} {\bibfnamefont {J.}~\bibnamefont
  {Rodr{\'i}guez-Carvajal}},\ }\href
  {http://www.sciencedirect.com/science/article/pii/092145269390108I}
  {\bibfield  {journal} {\bibinfo  {journal} {Physica B: Condensed Matter}\
  }\textbf {\bibinfo {volume} {192}},\ \bibinfo {pages} {55} (\bibinfo {year}
  {1993})}\BibitemShut {NoStop}%
\bibitem [{\citenamefont {Hohenberg}\ and\ \citenamefont
  {Kohn}(1964)}]{hohenberg1964}%
  \BibitemOpen
  \bibfield  {author} {\bibinfo {author} {\bibfnamefont {P.}~\bibnamefont
  {Hohenberg}}\ and\ \bibinfo {author} {\bibfnamefont {W.}~\bibnamefont
  {Kohn}},\ }\href {\doibase 10.1103/PhysRev.136.B864} {\bibfield  {journal}
  {\bibinfo  {journal} {Phys. Rev.}\ }\textbf {\bibinfo {volume} {136}},\
  \bibinfo {pages} {B864} (\bibinfo {year} {1964})}\BibitemShut {NoStop}%
\bibitem [{\citenamefont {Kohn}\ and\ \citenamefont {Sham}(1965)}]{kohn1965}%
  \BibitemOpen
  \bibfield  {author} {\bibinfo {author} {\bibfnamefont {W.}~\bibnamefont
  {Kohn}}\ and\ \bibinfo {author} {\bibfnamefont {L.~J.}\ \bibnamefont
  {Sham}},\ }\href {\doibase 10.1103/PhysRev.140.A1133} {\bibfield  {journal}
  {\bibinfo  {journal} {Phys. Rev.}\ }\textbf {\bibinfo {volume} {140}},\
  \bibinfo {pages} {A1133} (\bibinfo {year} {1965})}\BibitemShut {NoStop}%
\bibitem [{\citenamefont {Bl\"ochl}(1994)}]{blochl1994_PAW}%
  \BibitemOpen
  \bibfield  {author} {\bibinfo {author} {\bibfnamefont {P.~E.}\ \bibnamefont
  {Bl\"ochl}},\ }\href {\doibase 10.1103/PhysRevB.50.17953} {\bibfield
  {journal} {\bibinfo  {journal} {Phys. Rev. B}\ }\textbf {\bibinfo {volume}
  {50}},\ \bibinfo {pages} {17953} (\bibinfo {year} {1994})}\BibitemShut
  {NoStop}%
\bibitem [{\citenamefont {Kresse}\ and\ \citenamefont
  {Furthm\"uller}(1996)}]{kresse1996}%
  \BibitemOpen
  \bibfield  {author} {\bibinfo {author} {\bibfnamefont {G.}~\bibnamefont
  {Kresse}}\ and\ \bibinfo {author} {\bibfnamefont {J.}~\bibnamefont
  {Furthm\"uller}},\ }\href {\doibase 10.1103/PhysRevB.54.11169} {\bibfield
  {journal} {\bibinfo  {journal} {Phys. Rev. B}\ }\textbf {\bibinfo {volume}
  {54}},\ \bibinfo {pages} {11169} (\bibinfo {year} {1996})}\BibitemShut
  {NoStop}%
\bibitem [{\citenamefont {Kresse}\ and\ \citenamefont
  {Joubert}(1999)}]{kresse1999}%
  \BibitemOpen
  \bibfield  {author} {\bibinfo {author} {\bibfnamefont {G.}~\bibnamefont
  {Kresse}}\ and\ \bibinfo {author} {\bibfnamefont {D.}~\bibnamefont
  {Joubert}},\ }\href {\doibase 10.1103/PhysRevB.59.1758} {\bibfield  {journal}
  {\bibinfo  {journal} {Phys. Rev. B}\ }\textbf {\bibinfo {volume} {59}},\
  \bibinfo {pages} {1758} (\bibinfo {year} {1999})}\BibitemShut {NoStop}%
\bibitem [{\citenamefont {Perdew}\ \emph {et~al.}(1996)\citenamefont {Perdew},
  \citenamefont {Burke},\ and\ \citenamefont {Ernzerhof}}]{perdew1996_gga}%
  \BibitemOpen
  \bibfield  {author} {\bibinfo {author} {\bibfnamefont {J.~P.}\ \bibnamefont
  {Perdew}}, \bibinfo {author} {\bibfnamefont {K.}~\bibnamefont {Burke}}, \
  and\ \bibinfo {author} {\bibfnamefont {M.}~\bibnamefont {Ernzerhof}},\ }\href
  {\doibase 10.1103/PhysRevLett.77.3865} {\bibfield  {journal} {\bibinfo
  {journal} {Phys. Rev. Lett.}\ }\textbf {\bibinfo {volume} {77}},\ \bibinfo
  {pages} {3865} (\bibinfo {year} {1996})}\BibitemShut {NoStop}%
\bibitem [{\citenamefont {Fu}\ \emph {et~al.}(2007)\citenamefont {Fu},
  \citenamefont {Kane},\ and\ \citenamefont {Mele}}]{fu2007topological}%
  \BibitemOpen
  \bibfield  {author} {\bibinfo {author} {\bibfnamefont {L.}~\bibnamefont
  {Fu}}, \bibinfo {author} {\bibfnamefont {C.~L.}\ \bibnamefont {Kane}}, \ and\
  \bibinfo {author} {\bibfnamefont {E.~J.}\ \bibnamefont {Mele}},\ }\href@noop
  {} {\bibfield  {journal} {\bibinfo  {journal} {Physical review letters}\
  }\textbf {\bibinfo {volume} {98}},\ \bibinfo {pages} {106803} (\bibinfo
  {year} {2007})}\BibitemShut {NoStop}%
\bibitem [{\citenamefont {Lv}\ \emph {et~al.}(2015)\citenamefont {Lv},
  \citenamefont {Weng}, \citenamefont {Fu}, \citenamefont {Wang}, \citenamefont
  {Miao}, \citenamefont {Ma}, \citenamefont {Richard}, \citenamefont {Huang},
  \citenamefont {Zhao}, \citenamefont {Chen}, \citenamefont {Fang},
  \citenamefont {Dai}, \citenamefont {Qian},\ and\ \citenamefont
  {Ding}}]{TaAs_PRX}%
  \BibitemOpen
  \bibfield  {author} {\bibinfo {author} {\bibfnamefont {B.~Q.}\ \bibnamefont
  {Lv}}, \bibinfo {author} {\bibfnamefont {H.~M.}\ \bibnamefont {Weng}},
  \bibinfo {author} {\bibfnamefont {B.~B.}\ \bibnamefont {Fu}}, \bibinfo
  {author} {\bibfnamefont {X.~P.}\ \bibnamefont {Wang}}, \bibinfo {author}
  {\bibfnamefont {H.}~\bibnamefont {Miao}}, \bibinfo {author} {\bibfnamefont
  {J.}~\bibnamefont {Ma}}, \bibinfo {author} {\bibfnamefont {P.}~\bibnamefont
  {Richard}}, \bibinfo {author} {\bibfnamefont {X.~C.}\ \bibnamefont {Huang}},
  \bibinfo {author} {\bibfnamefont {L.~X.}\ \bibnamefont {Zhao}}, \bibinfo
  {author} {\bibfnamefont {G.~F.}\ \bibnamefont {Chen}}, \bibinfo {author}
  {\bibfnamefont {Z.}~\bibnamefont {Fang}}, \bibinfo {author} {\bibfnamefont
  {X.}~\bibnamefont {Dai}}, \bibinfo {author} {\bibfnamefont {T.}~\bibnamefont
  {Qian}}, \ and\ \bibinfo {author} {\bibfnamefont {H.}~\bibnamefont {Ding}},\
  }\href {\doibase 10.1103/PhysRevX.5.031013} {\bibfield  {journal} {\bibinfo
  {journal} {Phys. Rev. X}\ }\textbf {\bibinfo {volume} {5}},\ \bibinfo {pages}
  {031013} (\bibinfo {year} {2015})}\BibitemShut {NoStop}%
\bibitem [{\citenamefont {Hu}\ \emph {et~al.}(2017)\citenamefont {Hu},
  \citenamefont {Tang}, \citenamefont {Liu}, \citenamefont {Zhu}, \citenamefont
  {Wei},\ and\ \citenamefont {Mao}}]{NLSM_ZrSiS}%
  \BibitemOpen
  \bibfield  {author} {\bibinfo {author} {\bibfnamefont {J.}~\bibnamefont
  {Hu}}, \bibinfo {author} {\bibfnamefont {Z.}~\bibnamefont {Tang}}, \bibinfo
  {author} {\bibfnamefont {J.}~\bibnamefont {Liu}}, \bibinfo {author}
  {\bibfnamefont {Y.}~\bibnamefont {Zhu}}, \bibinfo {author} {\bibfnamefont
  {J.}~\bibnamefont {Wei}}, \ and\ \bibinfo {author} {\bibfnamefont
  {Z.}~\bibnamefont {Mao}},\ }\href {\doibase 10.1103/PhysRevB.96.045127}
  {\bibfield  {journal} {\bibinfo  {journal} {Phys. Rev. B}\ }\textbf {\bibinfo
  {volume} {96}},\ \bibinfo {pages} {045127} (\bibinfo {year}
  {2017})}\BibitemShut {NoStop}%
\bibitem [{\citenamefont {Hu}\ \emph {et~al.}(2016)\citenamefont {Hu},
  \citenamefont {Tang}, \citenamefont {Liu}, \citenamefont {Liu}, \citenamefont
  {Zhu}, \citenamefont {Graf}, \citenamefont {Myhro}, \citenamefont {Tran},
  \citenamefont {Lau}, \citenamefont {Wei},\ and\ \citenamefont
  {Mao}}]{NLSM_ZrSiSe_ZrSiTe}%
  \BibitemOpen
  \bibfield  {author} {\bibinfo {author} {\bibfnamefont {J.}~\bibnamefont
  {Hu}}, \bibinfo {author} {\bibfnamefont {Z.}~\bibnamefont {Tang}}, \bibinfo
  {author} {\bibfnamefont {J.}~\bibnamefont {Liu}}, \bibinfo {author}
  {\bibfnamefont {X.}~\bibnamefont {Liu}}, \bibinfo {author} {\bibfnamefont
  {Y.}~\bibnamefont {Zhu}}, \bibinfo {author} {\bibfnamefont {D.}~\bibnamefont
  {Graf}}, \bibinfo {author} {\bibfnamefont {K.}~\bibnamefont {Myhro}},
  \bibinfo {author} {\bibfnamefont {S.}~\bibnamefont {Tran}}, \bibinfo {author}
  {\bibfnamefont {C.~N.}\ \bibnamefont {Lau}}, \bibinfo {author} {\bibfnamefont
  {J.}~\bibnamefont {Wei}}, \ and\ \bibinfo {author} {\bibfnamefont
  {Z.}~\bibnamefont {Mao}},\ }\href {\doibase 10.1103/PhysRevLett.117.016602}
  {\bibfield  {journal} {\bibinfo  {journal} {Phys. Rev. Lett.}\ }\textbf
  {\bibinfo {volume} {117}},\ \bibinfo {pages} {016602} (\bibinfo {year}
  {2016})}\BibitemShut {NoStop}%
\bibitem [{\citenamefont {Du}\ \emph {et~al.}(2005)\citenamefont {Du},
  \citenamefont {Tsai}, \citenamefont {Maslov},\ and\ \citenamefont
  {Hebard}}]{PhysRevLett.94.166601}%
  \BibitemOpen
  \bibfield  {author} {\bibinfo {author} {\bibfnamefont {X.}~\bibnamefont
  {Du}}, \bibinfo {author} {\bibfnamefont {S.-W.}\ \bibnamefont {Tsai}},
  \bibinfo {author} {\bibfnamefont {D.~L.}\ \bibnamefont {Maslov}}, \ and\
  \bibinfo {author} {\bibfnamefont {A.~F.}\ \bibnamefont {Hebard}},\ }\href
  {\doibase 10.1103/PhysRevLett.94.166601} {\bibfield  {journal} {\bibinfo
  {journal} {Phys. Rev. Lett.}\ }\textbf {\bibinfo {volume} {94}},\ \bibinfo
  {pages} {166601} (\bibinfo {year} {2005})}\BibitemShut {NoStop}%
\bibitem [{\citenamefont {Tafti}\ \emph {et~al.}(2015)\citenamefont {Tafti},
  \citenamefont {Gibson}, \citenamefont {Kushwaha}, \citenamefont
  {Haldolaarachchige},\ and\ \citenamefont {Cava}}]{LaSb_NaturePhy_2015}%
  \BibitemOpen
  \bibfield  {author} {\bibinfo {author} {\bibfnamefont {F.~F.}\ \bibnamefont
  {Tafti}}, \bibinfo {author} {\bibfnamefont {Q.~D.}\ \bibnamefont {Gibson}},
  \bibinfo {author} {\bibfnamefont {S.~K.}\ \bibnamefont {Kushwaha}}, \bibinfo
  {author} {\bibfnamefont {N.}~\bibnamefont {Haldolaarachchige}}, \ and\
  \bibinfo {author} {\bibfnamefont {R.~J.}\ \bibnamefont {Cava}},\ }\href
  {https://doi.org/10.1038/nphys3581} {\bibfield  {journal} {\bibinfo
  {journal} {Nat. Phys.}\ }\textbf {\bibinfo {volume} {12}},\ \bibinfo {pages}
  {272} (\bibinfo {year} {2015})},\ \bibinfo {note} {article}\BibitemShut
  {NoStop}%
\bibitem [{\citenamefont {Li}\ \emph {et~al.}(2016)\citenamefont {Li},
  \citenamefont {Li}, \citenamefont {Wang}, \citenamefont {Wang}, \citenamefont
  {Xu}, \citenamefont {Xi}, \citenamefont {Cao},\ and\ \citenamefont
  {Dai}}]{TaSb2_PRB(R)_2016}%
  \BibitemOpen
  \bibfield  {author} {\bibinfo {author} {\bibfnamefont {Y.}~\bibnamefont
  {Li}}, \bibinfo {author} {\bibfnamefont {L.}~\bibnamefont {Li}}, \bibinfo
  {author} {\bibfnamefont {J.}~\bibnamefont {Wang}}, \bibinfo {author}
  {\bibfnamefont {T.}~\bibnamefont {Wang}}, \bibinfo {author} {\bibfnamefont
  {X.}~\bibnamefont {Xu}}, \bibinfo {author} {\bibfnamefont {C.}~\bibnamefont
  {Xi}}, \bibinfo {author} {\bibfnamefont {C.}~\bibnamefont {Cao}}, \ and\
  \bibinfo {author} {\bibfnamefont {J.}~\bibnamefont {Dai}},\ }\href {\doibase
  10.1103/PhysRevB.94.121115} {\bibfield  {journal} {\bibinfo  {journal} {Phys.
  Rev. B}\ }\textbf {\bibinfo {volume} {94}},\ \bibinfo {pages} {121115}
  (\bibinfo {year} {2016})}\BibitemShut {NoStop}%
\bibitem [{\citenamefont {Zeng}\ \emph {et~al.}(2016)\citenamefont {Zeng},
  \citenamefont {Lou}, \citenamefont {Wu}, \citenamefont {Xu}, \citenamefont
  {Guo}, \citenamefont {Kong}, \citenamefont {Zhong}, \citenamefont {Ma},
  \citenamefont {Fu}, \citenamefont {Richard}, \citenamefont {Wang},
  \citenamefont {Liu}, \citenamefont {Lu}, \citenamefont {Huang}, \citenamefont
  {Fang}, \citenamefont {Sun}, \citenamefont {Wang}, \citenamefont {Wang},
  \citenamefont {Shi}, \citenamefont {Weng}, \citenamefont {Lei}, \citenamefont
  {Liu}, \citenamefont {Wang}, \citenamefont {Qian}, \citenamefont {Luo},\ and\
  \citenamefont {Ding}}]{LaSb_PRL_2016}%
  \BibitemOpen
  \bibfield  {author} {\bibinfo {author} {\bibfnamefont {L.-K.}\ \bibnamefont
  {Zeng}}, \bibinfo {author} {\bibfnamefont {R.}~\bibnamefont {Lou}}, \bibinfo
  {author} {\bibfnamefont {D.-S.}\ \bibnamefont {Wu}}, \bibinfo {author}
  {\bibfnamefont {Q.~N.}\ \bibnamefont {Xu}}, \bibinfo {author} {\bibfnamefont
  {P.-J.}\ \bibnamefont {Guo}}, \bibinfo {author} {\bibfnamefont {L.-Y.}\
  \bibnamefont {Kong}}, \bibinfo {author} {\bibfnamefont {Y.-G.}\ \bibnamefont
  {Zhong}}, \bibinfo {author} {\bibfnamefont {J.-Z.}\ \bibnamefont {Ma}},
  \bibinfo {author} {\bibfnamefont {B.-B.}\ \bibnamefont {Fu}}, \bibinfo
  {author} {\bibfnamefont {P.}~\bibnamefont {Richard}}, \bibinfo {author}
  {\bibfnamefont {P.}~\bibnamefont {Wang}}, \bibinfo {author} {\bibfnamefont
  {G.~T.}\ \bibnamefont {Liu}}, \bibinfo {author} {\bibfnamefont
  {L.}~\bibnamefont {Lu}}, \bibinfo {author} {\bibfnamefont {Y.-B.}\
  \bibnamefont {Huang}}, \bibinfo {author} {\bibfnamefont {C.}~\bibnamefont
  {Fang}}, \bibinfo {author} {\bibfnamefont {S.-S.}\ \bibnamefont {Sun}},
  \bibinfo {author} {\bibfnamefont {Q.}~\bibnamefont {Wang}}, \bibinfo {author}
  {\bibfnamefont {L.}~\bibnamefont {Wang}}, \bibinfo {author} {\bibfnamefont
  {Y.-G.}\ \bibnamefont {Shi}}, \bibinfo {author} {\bibfnamefont {H.~M.}\
  \bibnamefont {Weng}}, \bibinfo {author} {\bibfnamefont {H.-C.}\ \bibnamefont
  {Lei}}, \bibinfo {author} {\bibfnamefont {K.}~\bibnamefont {Liu}}, \bibinfo
  {author} {\bibfnamefont {S.-C.}\ \bibnamefont {Wang}}, \bibinfo {author}
  {\bibfnamefont {T.}~\bibnamefont {Qian}}, \bibinfo {author} {\bibfnamefont
  {J.-L.}\ \bibnamefont {Luo}}, \ and\ \bibinfo {author} {\bibfnamefont
  {H.}~\bibnamefont {Ding}},\ }\href {\doibase 10.1103/PhysRevLett.117.127204}
  {\bibfield  {journal} {\bibinfo  {journal} {Phys. Rev. Lett.}\ }\textbf
  {\bibinfo {volume} {117}},\ \bibinfo {pages} {127204} (\bibinfo {year}
  {2016})}\BibitemShut {NoStop}%
\bibitem [{\citenamefont {Singha}\ \emph
  {et~al.}(2017{\natexlab{b}})\citenamefont {Singha}, \citenamefont {Pariari},
  \citenamefont {Satpati},\ and\ \citenamefont {Mandal}}]{LaSbTe_pPRB_2017}%
  \BibitemOpen
  \bibfield  {author} {\bibinfo {author} {\bibfnamefont {R.}~\bibnamefont
  {Singha}}, \bibinfo {author} {\bibfnamefont {A.}~\bibnamefont {Pariari}},
  \bibinfo {author} {\bibfnamefont {B.}~\bibnamefont {Satpati}}, \ and\
  \bibinfo {author} {\bibfnamefont {P.}~\bibnamefont {Mandal}},\ }\href
  {\doibase 10.1103/PhysRevB.96.245138} {\bibfield  {journal} {\bibinfo
  {journal} {Phys. Rev. B}\ }\textbf {\bibinfo {volume} {96}},\ \bibinfo
  {pages} {245138} (\bibinfo {year} {2017}{\natexlab{b}})}\BibitemShut
  {NoStop}%
\bibitem [{\citenamefont {Du}\ \emph {et~al.}(2018)\citenamefont {Du},
  \citenamefont {Lou}, \citenamefont {Zhang}, \citenamefont {Zhou},
  \citenamefont {Xu}, \citenamefont {Chen}, \citenamefont {Tang}, \citenamefont
  {Chen}, \citenamefont {Chen}, \citenamefont {Zhu}, \citenamefont {Wang},
  \citenamefont {Yang}, \citenamefont {Wu}, \citenamefont {Yazyev},\ and\
  \citenamefont {Fang}}]{du2018}%
  \BibitemOpen
  \bibfield  {author} {\bibinfo {author} {\bibfnamefont {J.}~\bibnamefont
  {Du}}, \bibinfo {author} {\bibfnamefont {Z.}~\bibnamefont {Lou}}, \bibinfo
  {author} {\bibfnamefont {S.}~\bibnamefont {Zhang}}, \bibinfo {author}
  {\bibfnamefont {Y.}~\bibnamefont {Zhou}}, \bibinfo {author} {\bibfnamefont
  {B.}~\bibnamefont {Xu}}, \bibinfo {author} {\bibfnamefont {Q.}~\bibnamefont
  {Chen}}, \bibinfo {author} {\bibfnamefont {Y.}~\bibnamefont {Tang}}, \bibinfo
  {author} {\bibfnamefont {S.}~\bibnamefont {Chen}}, \bibinfo {author}
  {\bibfnamefont {H.}~\bibnamefont {Chen}}, \bibinfo {author} {\bibfnamefont
  {Q.}~\bibnamefont {Zhu}}, \bibinfo {author} {\bibfnamefont {H.}~\bibnamefont
  {Wang}}, \bibinfo {author} {\bibfnamefont {J.}~\bibnamefont {Yang}}, \bibinfo
  {author} {\bibfnamefont {Q.}~\bibnamefont {Wu}}, \bibinfo {author}
  {\bibfnamefont {O.~V.}\ \bibnamefont {Yazyev}}, \ and\ \bibinfo {author}
  {\bibfnamefont {M.}~\bibnamefont {Fang}},\ }\href {\doibase
  10.1103/PhysRevB.97.245101} {\bibfield  {journal} {\bibinfo  {journal} {Phys.
  Rev. B}\ }\textbf {\bibinfo {volume} {97}},\ \bibinfo {pages} {245101}
  (\bibinfo {year} {2018})}\BibitemShut {NoStop}%
\bibitem [{\citenamefont {Zhao}\ \emph {et~al.}(2015)\citenamefont {Zhao},
  \citenamefont {Liu}, \citenamefont {Yan}, \citenamefont {An}, \citenamefont
  {Liu}, \citenamefont {Zhang}, \citenamefont {Wang}, \citenamefont {Liu},
  \citenamefont {Jiang}, \citenamefont {Li}, \citenamefont {Wang},
  \citenamefont {Li}, \citenamefont {Mandrus}, \citenamefont {Xie},
  \citenamefont {Pan},\ and\ \citenamefont {Wang}}]{WTe2_PRB_R_2015}%
  \BibitemOpen
  \bibfield  {author} {\bibinfo {author} {\bibfnamefont {Y.}~\bibnamefont
  {Zhao}}, \bibinfo {author} {\bibfnamefont {H.}~\bibnamefont {Liu}}, \bibinfo
  {author} {\bibfnamefont {J.}~\bibnamefont {Yan}}, \bibinfo {author}
  {\bibfnamefont {W.}~\bibnamefont {An}}, \bibinfo {author} {\bibfnamefont
  {J.}~\bibnamefont {Liu}}, \bibinfo {author} {\bibfnamefont {X.}~\bibnamefont
  {Zhang}}, \bibinfo {author} {\bibfnamefont {H.}~\bibnamefont {Wang}},
  \bibinfo {author} {\bibfnamefont {Y.}~\bibnamefont {Liu}}, \bibinfo {author}
  {\bibfnamefont {H.}~\bibnamefont {Jiang}}, \bibinfo {author} {\bibfnamefont
  {Q.}~\bibnamefont {Li}}, \bibinfo {author} {\bibfnamefont {Y.}~\bibnamefont
  {Wang}}, \bibinfo {author} {\bibfnamefont {X.-Z.}\ \bibnamefont {Li}},
  \bibinfo {author} {\bibfnamefont {D.}~\bibnamefont {Mandrus}}, \bibinfo
  {author} {\bibfnamefont {X.~C.}\ \bibnamefont {Xie}}, \bibinfo {author}
  {\bibfnamefont {M.}~\bibnamefont {Pan}}, \ and\ \bibinfo {author}
  {\bibfnamefont {J.}~\bibnamefont {Wang}},\ }\href {\doibase
  10.1103/PhysRevB.92.041104} {\bibfield  {journal} {\bibinfo  {journal} {Phys.
  Rev. B}\ }\textbf {\bibinfo {volume} {92}},\ \bibinfo {pages} {041104}
  (\bibinfo {year} {2015})}\BibitemShut {NoStop}%
\bibitem [{\citenamefont {Wang}\ \emph
  {et~al.}(2017{\natexlab{b}})\citenamefont {Wang}, \citenamefont {Graf},
  \citenamefont {Liu}, \citenamefont {Du}, \citenamefont {Zheng}, \citenamefont
  {Lei},\ and\ \citenamefont {Petrovic}}]{WP_PRB_2017}%
  \BibitemOpen
  \bibfield  {author} {\bibinfo {author} {\bibfnamefont {A.}~\bibnamefont
  {Wang}}, \bibinfo {author} {\bibfnamefont {D.}~\bibnamefont {Graf}}, \bibinfo
  {author} {\bibfnamefont {Y.}~\bibnamefont {Liu}}, \bibinfo {author}
  {\bibfnamefont {Q.}~\bibnamefont {Du}}, \bibinfo {author} {\bibfnamefont
  {J.}~\bibnamefont {Zheng}}, \bibinfo {author} {\bibfnamefont
  {H.}~\bibnamefont {Lei}}, \ and\ \bibinfo {author} {\bibfnamefont
  {C.}~\bibnamefont {Petrovic}},\ }\href {\doibase 10.1103/PhysRevB.96.121107}
  {\bibfield  {journal} {\bibinfo  {journal} {Phys. Rev. B}\ }\textbf {\bibinfo
  {volume} {96}},\ \bibinfo {pages} {121107} (\bibinfo {year}
  {2017}{\natexlab{b}})}\BibitemShut {NoStop}%
\bibitem [{\citenamefont {Xu}\ \emph {et~al.}(2015)\citenamefont {Xu},
  \citenamefont {Jiao}, \citenamefont {Zhou}, \citenamefont {Guo},
  \citenamefont {Li}, \citenamefont {Dai}, \citenamefont {Lin}, \citenamefont
  {Liu}, \citenamefont {Zhu}, \citenamefont {Lu}, \citenamefont {Yuan},\ and\
  \citenamefont {Cao}}]{Xu_2015}%
  \BibitemOpen
  \bibfield  {author} {\bibinfo {author} {\bibfnamefont {X.}~\bibnamefont
  {Xu}}, \bibinfo {author} {\bibfnamefont {W.~H.}\ \bibnamefont {Jiao}},
  \bibinfo {author} {\bibfnamefont {N.}~\bibnamefont {Zhou}}, \bibinfo {author}
  {\bibfnamefont {Y.}~\bibnamefont {Guo}}, \bibinfo {author} {\bibfnamefont
  {Y.~K.}\ \bibnamefont {Li}}, \bibinfo {author} {\bibfnamefont
  {J.}~\bibnamefont {Dai}}, \bibinfo {author} {\bibfnamefont {Z.~Q.}\
  \bibnamefont {Lin}}, \bibinfo {author} {\bibfnamefont {Y.~J.}\ \bibnamefont
  {Liu}}, \bibinfo {author} {\bibfnamefont {Z.}~\bibnamefont {Zhu}}, \bibinfo
  {author} {\bibfnamefont {X.}~\bibnamefont {Lu}}, \bibinfo {author}
  {\bibfnamefont {H.~Q.}\ \bibnamefont {Yuan}}, \ and\ \bibinfo {author}
  {\bibfnamefont {G.}~\bibnamefont {Cao}},\ }\href {\doibase
  10.1088/0953-8984/27/33/335701} {\bibfield  {journal} {\bibinfo  {journal}
  {Journal of Physics: Condensed Matter}\ }\textbf {\bibinfo {volume} {27}},\
  \bibinfo {pages} {335701} (\bibinfo {year} {2015})}\BibitemShut {NoStop}%
\bibitem [{\citenamefont {Sun}\ \emph {et~al.}(2014)\citenamefont {Sun},
  \citenamefont {Taen}, \citenamefont {Yamada}, \citenamefont {Pyon},
  \citenamefont {Nishizaki}, \citenamefont {Shi},\ and\ \citenamefont
  {Tamegai}}]{Multiband_Kohler}%
  \BibitemOpen
  \bibfield  {author} {\bibinfo {author} {\bibfnamefont {Y.}~\bibnamefont
  {Sun}}, \bibinfo {author} {\bibfnamefont {T.}~\bibnamefont {Taen}}, \bibinfo
  {author} {\bibfnamefont {T.}~\bibnamefont {Yamada}}, \bibinfo {author}
  {\bibfnamefont {S.}~\bibnamefont {Pyon}}, \bibinfo {author} {\bibfnamefont
  {T.}~\bibnamefont {Nishizaki}}, \bibinfo {author} {\bibfnamefont
  {Z.}~\bibnamefont {Shi}}, \ and\ \bibinfo {author} {\bibfnamefont
  {T.}~\bibnamefont {Tamegai}},\ }\href {\doibase 10.1103/PhysRevB.89.144512}
  {\bibfield  {journal} {\bibinfo  {journal} {Phys. Rev. B}\ }\textbf {\bibinfo
  {volume} {89}},\ \bibinfo {pages} {144512} (\bibinfo {year}
  {2014})}\BibitemShut {NoStop}%
\bibitem [{\citenamefont {McKenzie}\ \emph {et~al.}(1998)\citenamefont
  {McKenzie}, \citenamefont {Qualls}, \citenamefont {Han},\ and\ \citenamefont
  {Brooks}}]{McKenzie_Kohler}%
  \BibitemOpen
  \bibfield  {author} {\bibinfo {author} {\bibfnamefont {R.~H.}\ \bibnamefont
  {McKenzie}}, \bibinfo {author} {\bibfnamefont {J.~S.}\ \bibnamefont
  {Qualls}}, \bibinfo {author} {\bibfnamefont {S.~Y.}\ \bibnamefont {Han}}, \
  and\ \bibinfo {author} {\bibfnamefont {J.~S.}\ \bibnamefont {Brooks}},\
  }\href {\doibase 10.1103/PhysRevB.57.11854} {\bibfield  {journal} {\bibinfo
  {journal} {Phys. Rev. B}\ }\textbf {\bibinfo {volume} {57}},\ \bibinfo
  {pages} {11854} (\bibinfo {year} {1998})}\BibitemShut {NoStop}%
\bibitem [{\citenamefont {Xu}\ \emph {et~al.}(1997)\citenamefont {Xu},
  \citenamefont {Husmann}, \citenamefont {Rosenbaum}, \citenamefont {Saboungi},
  \citenamefont {Enderby},\ and\ \citenamefont
  {Littlewood}}]{Ag2Se_Nature_1997}%
  \BibitemOpen
  \bibfield  {author} {\bibinfo {author} {\bibfnamefont {R.}~\bibnamefont
  {Xu}}, \bibinfo {author} {\bibfnamefont {A.}~\bibnamefont {Husmann}},
  \bibinfo {author} {\bibfnamefont {T.~F.}\ \bibnamefont {Rosenbaum}}, \bibinfo
  {author} {\bibfnamefont {M.-L.}\ \bibnamefont {Saboungi}}, \bibinfo {author}
  {\bibfnamefont {J.~E.}\ \bibnamefont {Enderby}}, \ and\ \bibinfo {author}
  {\bibfnamefont {P.~B.}\ \bibnamefont {Littlewood}},\ }\href {\doibase
  10.1038/36306} {\bibfield  {journal} {\bibinfo  {journal} {Nature}\ }\textbf
  {\bibinfo {volume} {390}},\ \bibinfo {pages} {57} (\bibinfo {year}
  {1997})}\BibitemShut {NoStop}%
\bibitem [{\citenamefont {Kozlova}\ \emph {et~al.}(2012)\citenamefont
  {Kozlova}, \citenamefont {Mori}, \citenamefont {Makarovsky}, \citenamefont
  {Eaves}, \citenamefont {Zhuang}, \citenamefont {Krier},\ and\ \citenamefont
  {Patan{\`e}}}]{Kozlova_2012}%
  \BibitemOpen
  \bibfield  {author} {\bibinfo {author} {\bibfnamefont {N.~V.}\ \bibnamefont
  {Kozlova}}, \bibinfo {author} {\bibfnamefont {N.}~\bibnamefont {Mori}},
  \bibinfo {author} {\bibfnamefont {O.}~\bibnamefont {Makarovsky}}, \bibinfo
  {author} {\bibfnamefont {L.}~\bibnamefont {Eaves}}, \bibinfo {author}
  {\bibfnamefont {Q.~D.}\ \bibnamefont {Zhuang}}, \bibinfo {author}
  {\bibfnamefont {A.}~\bibnamefont {Krier}}, \ and\ \bibinfo {author}
  {\bibfnamefont {A.}~\bibnamefont {Patan{\`e}}},\ }\href {\doibase
  10.1038/ncomms2106} {\bibfield  {journal} {\bibinfo  {journal} {Nature
  Communications}\ }\textbf {\bibinfo {volume} {3}},\ \bibinfo {pages} {1097}
  (\bibinfo {year} {2012})}\BibitemShut {NoStop}%
\bibitem [{\citenamefont {Wu}\ \emph {et~al.}(2020)\citenamefont {Wu},
  \citenamefont {Barrena}, \citenamefont {Suderow},\ and\ \citenamefont
  {Guillamón}}]{PtBi2_arxiv_2020}%
  \BibitemOpen
  \bibfield  {author} {\bibinfo {author} {\bibfnamefont {B.}~\bibnamefont
  {Wu}}, \bibinfo {author} {\bibfnamefont {V.}~\bibnamefont {Barrena}},
  \bibinfo {author} {\bibfnamefont {H.}~\bibnamefont {Suderow}}, \ and\
  \bibinfo {author} {\bibfnamefont {I.}~\bibnamefont {Guillamón}},\
  }\href@noop {} {\enquote {\bibinfo {title} {Huge linear magnetoresistance due
  to open orbits in $γ$-ptbi$_2$},}\ } (\bibinfo {year} {2020}),\ \Eprint
  {http://arxiv.org/abs/2001.07674} {arXiv:2001.07674 [cond-mat.mtrl-sci]}
  \BibitemShut {NoStop}%
\bibitem [{\citenamefont {Feng}\ \emph {et~al.}(2019)\citenamefont {Feng},
  \citenamefont {Wang}, \citenamefont {Silevitch}, \citenamefont {Yan},
  \citenamefont {Kobayashi}, \citenamefont {Hedo}, \citenamefont {Nakama},
  \citenamefont {Onuki}, \citenamefont {Suslov}, \citenamefont {Mihaila},
  \citenamefont {Littlewood},\ and\ \citenamefont {Rosenbaum}}]{Feng11201}%
  \BibitemOpen
  \bibfield  {author} {\bibinfo {author} {\bibfnamefont {Y.}~\bibnamefont
  {Feng}}, \bibinfo {author} {\bibfnamefont {Y.}~\bibnamefont {Wang}}, \bibinfo
  {author} {\bibfnamefont {D.~M.}\ \bibnamefont {Silevitch}}, \bibinfo {author}
  {\bibfnamefont {J.-Q.}\ \bibnamefont {Yan}}, \bibinfo {author} {\bibfnamefont
  {R.}~\bibnamefont {Kobayashi}}, \bibinfo {author} {\bibfnamefont
  {M.}~\bibnamefont {Hedo}}, \bibinfo {author} {\bibfnamefont {T.}~\bibnamefont
  {Nakama}}, \bibinfo {author} {\bibfnamefont {Y.}~\bibnamefont {Onuki}},
  \bibinfo {author} {\bibfnamefont {A.~V.}\ \bibnamefont {Suslov}}, \bibinfo
  {author} {\bibfnamefont {B.}~\bibnamefont {Mihaila}}, \bibinfo {author}
  {\bibfnamefont {P.~B.}\ \bibnamefont {Littlewood}}, \ and\ \bibinfo {author}
  {\bibfnamefont {T.~F.}\ \bibnamefont {Rosenbaum}},\ }\href {\doibase
  10.1073/pnas.1820092116} {\bibfield  {journal} {\bibinfo  {journal}
  {Proceedings of the National Academy of Sciences}\ }\textbf {\bibinfo
  {volume} {116}},\ \bibinfo {pages} {11201} (\bibinfo {year}
  {2019})}\BibitemShut {NoStop}%
\bibitem [{\citenamefont {Liang}\ \emph {et~al.}(2014)\citenamefont {Liang},
  \citenamefont {Gibson}, \citenamefont {Ali}, \citenamefont {Liu},
  \citenamefont {Cava},\ and\ \citenamefont {Ong}}]{Cd3AS2_nature_2015}%
  \BibitemOpen
  \bibfield  {author} {\bibinfo {author} {\bibfnamefont {T.}~\bibnamefont
  {Liang}}, \bibinfo {author} {\bibfnamefont {Q.}~\bibnamefont {Gibson}},
  \bibinfo {author} {\bibfnamefont {M.~N.}\ \bibnamefont {Ali}}, \bibinfo
  {author} {\bibfnamefont {M.}~\bibnamefont {Liu}}, \bibinfo {author}
  {\bibfnamefont {R.~J.}\ \bibnamefont {Cava}}, \ and\ \bibinfo {author}
  {\bibfnamefont {N.~P.}\ \bibnamefont {Ong}},\ }\href
  {https://doi.org/10.1038/nmat4143} {\bibfield  {journal} {\bibinfo  {journal}
  {Nature Materials}\ }\textbf {\bibinfo {volume} {14}},\ \bibinfo {pages} {280
  EP } (\bibinfo {year} {2014})}\BibitemShut {NoStop}%
\bibitem [{\citenamefont {Tang}\ \emph {et~al.}(2011)\citenamefont {Tang},
  \citenamefont {Liang}, \citenamefont {Qiu},\ and\ \citenamefont
  {Gao}}]{Bi2Se3_TIs}%
  \BibitemOpen
  \bibfield  {author} {\bibinfo {author} {\bibfnamefont {H.}~\bibnamefont
  {Tang}}, \bibinfo {author} {\bibfnamefont {D.}~\bibnamefont {Liang}},
  \bibinfo {author} {\bibfnamefont {R.~L.~J.}\ \bibnamefont {Qiu}}, \ and\
  \bibinfo {author} {\bibfnamefont {X.~P.~A.}\ \bibnamefont {Gao}},\
  }\href@noop {} {\bibfield  {journal} {\bibinfo  {journal} {ACS nano}\
  }\textbf {\bibinfo {volume} {5 9}},\ \bibinfo {pages} {7510} (\bibinfo {year}
  {2011})}\BibitemShut {NoStop}%
\end{thebibliography}%

\end{document}